\documentclass[12pt,a4paper]{elsarticle}

\usepackage{cancel}

\usepackage{lineno,hyperref}
\modulolinenumbers[5]

\journal{International Communications in Heat and Mass Transfer}

\biboptions{sort&compress}
\usepackage[T1]{fontenc}
\usepackage{amsmath,amssymb,amsthm,amsfonts}
\usepackage{siunitx}	
\usepackage{subcaption}
\usepackage{float}
\usepackage{verbatim}
\usepackage[font=small]{caption}
\usepackage{multirow}

\usepackage{xcolor}
\usepackage{blindtext}
\usepackage{multirow,multicol,colortbl}
\usepackage{tikz}

\def\p{\partial}

\def\({\text{\huge (}}
\def\){\text{\huge )}}

\def\]{\text{\huge ]}}
\def\[{\text{\huge [}}

\newcommand{\ud}{\mathop{}\!\mathrm{d}}

\newcommand{\bi}{\begin{itemize}}
\newcommand{\ei}{\end{itemize}}
\newcommand{\be}{\begin{equation}}
\newcommand{\ee}{\end{equation}}
\newcommand{\ba}{\begin{align}}
\newcommand{\ea}{\end{align}}

\newcommand\nc{\newcommand}
\nc\pad[2]{\frac{\p #1}{\p #2}} \nc\padd[2]{\frac{\p^2 #1}{\p
{#2}^2}} \nc\nd[2]{\frac{\ud #1}{\ud #2}} \nc\pat[2]{\frac{D #1}{D
#2}} \nc\ov{\overline} \nc\degree{^{\circ}} \nc\ord[1]{{\cal
O}(#1)} \nc\ra{\rightarrow} \nc\Ra{\Rightarrow} \nc\dint{{\mbox ~
d}}

\newcommand{\bea}{\begin{eqnarray}}
\newcommand{\eea}{\end{eqnarray}}
\newcommand{\beas}{\begin{eqnarray*}}
\newcommand{\eeas}{\end{eqnarray*}}

\begin{document}

\begin{frontmatter}

\title{
Modelling large mass removal in adsorption columns
}

\author{T. G. Myers}
\address{Centre de Recerca Matem\`atica, Edifici C, Campus Bellaterra, 08193 Bellaterra, Spain}
\author{M. Calvo-Schwarzwalder}
\address{Centre de Recerca Matem\`atica, Edifici C, Campus Bellaterra, 08193 Bellaterra, Spain}
\author{F. Font\fnref{myfootnote}}
\address{Department of Fluid Mechanics, Universitat Polit\`ecnica de Catalunya, 08019 Barcelona, Spain}
\author{A. Valverde}
\address{Department of Chemical Engineering, Universitat Polit\`ecnica de Catalunya, 08028 Barcelona, Spain.}
\fntext[myfootnote]{Corresponding author: francesc.font@upc.edu}

\begin{abstract} 
A mathematical model is developed to describe column adsorption when the contaminant constitutes a significant amount of the fluid. This requires modelling the variation of pressure and velocity, in addition to the usual advection-diffusion-adsorption and kinetic equations describing concentration and adsorption rates. The model builds on previous work based on a linear kinetic equation, to include both physical and chemical adsorption. A semi-analytical solution is developed and validated against a numerical solution. The model is tested against experimental data for the adsorption of large quantities of CO$_2$ from a helium mixture, with a CO$_2$ volume fraction ranging from 14\% to 69\%, and a N$_2$ mixture with 16\% to 33\% CO$_2$ volume fraction. Our results show a significant improvement with respect to models for the removal of trace amounts of contaminant.
\end{abstract}

\begin{keyword}
Contaminant removal \sep Pollutant removal \sep Adsorption \sep Fluid dynamics \sep Mathematical model 
\MSC[2010] 35Q35
\end{keyword}

\end{frontmatter}

%%%%%%%%%%%%%%%%%%%%%%%%%%%%%%%%%%%%%%%%%%%%%%%%%%%%%%%%%%%%%%%%%%%%%%%%%%%%%%%%%%%%%%%%

\section{Introduction}\label{sec:intro}

Column sorption is a practical method for removing a contaminant from a carrier fluid. It has uses in environmental applications such as greenhouse gas capture, groundwater remediation and biogas cleansing as well as industrial uses such as the purification of biopharmaceutical products, the cleansing of flue gases, biofuel purification and many more. Given the importance of the process, particularly in environmental remediation, researchers constantly strive to improve and optimise the technology.

A key element in the optimisation process is the development and understanding of models for adsorption. In general 
mathematical models have focussed on the removal of trace amounts of contaminants subject to a physisorption process. The base model for this situation results in a system of two equations, describing the evolution of the cross-sectionally averaged contaminant concentration and amount adsorbed through a column. Perhaps the most well-known solution to the system is that of Bohart and Adams \cite{Bohart} which predicts the concentration throughout the column and provides a simple expression for the breakthrough curve (the outlet concentration). Their model is frequently referred to as the Thomas model, with a minor rearrangement it is also referred to as the Bed Depth Service Time Model. The Bohart-Adams model is based on an adsorption rate similar to the standard Langmuir model but with zero desorption. Although it can match certain data sets with others it shows very poor agreement. Amundsen \cite{Amund} attempted to improve agreement by shifting the time axis slightly (by the column length divided by the interstitial fluid velocity). In practice this makes a negligible difference.  The Yoon-Nelson model \cite{Yoon} takes the same mathematical form as Bohart-Adams but is based on the probability of a molecule escaping the column outlet. In addition to these theoretically based models there exist many empirical ones. Shafeeyan {\it et al} \cite{Shaf14} review over “three decades” of modelling: all models follow the same format and are solved numerically. They go on to state that since the computational time is so large it would be desirable to develop reduced models with good prediction capabilities and so facilitate optimisation. Li \textit{et al} \cite{Li} state that "modelling methods with mathematical equations are still rare in the existing publications". 
A discussion of errors and inaccuracies of breakthrough models is provided in \cite{Myers24}.

When dealing with the removal of a significant quantity of material a model should account for the effect of the mass loss, for example on velocity and pressure. However, it is common practice to employ the Bohart-Adams form to fit breakthrough curves with a 15\% v/v or higher content of contaminant \cite{Chiang2017,Huang2019,Monazam2013}, or determine numerical solutions of single-component constant velocity models \cite{Moreira2006,Sabouni2013,Shafeeyan2015}. Certain studies differentiate between the fluid components but still impose constant velocity \cite{BastosNeto2011}. In studies where the effect of the removal of large quantities of contaminant is accounted for (20\% up to 75\% v/v, for instance) \cite{Delgado2006,Ding2000,Casas2012},  a numerical solution is applied  using the Ergun equation \cite{Ergun1949} to relate the velocity with the pressure drop, without considering the sink term in the conservation of momentum. Dantas \textit{et al } \cite{Dantas,Dantas2} employ the Ergun equation, a pseudo-first order kinetic equation (also known as the linear driving force model) and include temperature effects. Their system requires a full numerical solution. 

In a series of papers our group has explained why the Bohart-Adams and related models may be inaccurate \cite{Myers19,Myers23,Myers24} and then produced accurate models to describe chemisorption \cite{Aguareles23},  intra-particle diffusion \cite{Valverde2024,Auton2024} and  extraction \cite{myers2022}. However, all of these models are designed to deal with the removal of trace amounts of contaminants and so are not suitable to deal with large emissions, such as flue  or exhaust gases.

In \cite{Myers20} a model was developed and analysed to describe the capture of significant quantities of CO$_2$, where the adsorption was also described by the linear driving force model. Although mathematically convenient, this form presents physical issues in that adsorption occurs independently of whether there is a contaminant in the fluid or not. It also neglects desorption. Valverde \textit{et al} \cite{Valverde2024} demonstrated that this type of kinetic equation can predict negative concentrations near the contaminant front. 
 
In the present work we will extend the large mass removal model of \cite{Myers20}, applying a physically realistic kinetic equation which encompasses both physical and chemical adsorption. 
In \S\ref{sec:model} we discuss the experimental setup and derive the general mathematical model. After introducing appropriate non-dimensional variables in  \S\ref{sec:simplification} it is demonstrated that the model may be simplified without losing accuracy. In \S\ref{sec:SolutionMethods} we present analytical solutions to the reduced model  and verify them by comparison with  a numerical solution. Results of the model are then presented, demonstrating the effect of key parameters, such as contaminant volume fraction, system temperature, or flow rate. In \S\ref{sec:AnalyticsVSExperiments} we discuss practical applications of the model and compare the analytical solutions against experimental data for the removal of CO$_2$ from a CO$_2$/N$_2$ mixture. We also discuss an alternative way to present the data which makes the correct model easier to identify.

\section{Mathematical Model}\label{sec:model}
In this section we first present the experimental set up that motivates the mathematical model and then discuss the model in the context of removal of large amounts of contaminant.

\subsection{Description of the experimental set up and key variables}
\begin{figure}[h!]
    \centering
    \includegraphics{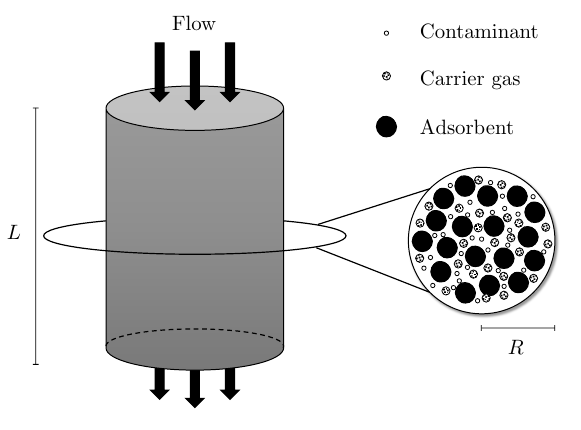}
    % \begin{tikzpicture}
    %     \draw (0, 0) node[inner sep=0] {\includegraphics[width=.6\textwidth]{LargeMassScheme.pdf}};
    %     \draw (-1.8, 3.5) node {Flow};
    %     \draw (-5, -.4) node {$L$};
    %     \draw (4, -2.5) node {$R$};
    %     \draw[anchor=west] (2, 1.55) node {Adsorbent};
    %     \draw[anchor=west] (2, 2.425) node {Carrier gas};
    %     \draw[anchor=west] (2, 3.3) node {Contaminant};
    % \end{tikzpicture} 
    \caption{Schematic of the experimental setup. We consider a column of length $L$ and radius $R$ that is packed with a porous, adsorbent material. A gas mixture consisting of a carrier gas and a contaminant is forced into the column at the inlet and the concentrations at the outlet are measured throughout the experiment.}
    \label{fig:scheme}
\end{figure}

As depicted in Fig. \ref{fig:scheme}, we consider a column of length $L$ and radius $R$  filled with a porous material,  the absorbent, that occupies a fraction $1-\epsilon$ of the total space available. At the inlet, we introduce a fluid mixture with fixed contaminant and carrier gas volume fractions $\phi_1$ and $\phi_2$ (with $\phi_1+\phi_2=1$) at a constant velocity $u_0$. The contaminant attaches to the adsorbent's surface as the mixture flows through the column.  The concentration of contaminant is measured at the outlet to record the \textit{breakthrough curve}. To model this process, at a minimum, we must calculate the concentration of both carrier fluid and contaminant, the amount adsorbed, pressure and velocity fields. 
In  Myers \emph{et al.} \cite{Myers19} temperature variation was also included in the initial system, but later shown to be negligible. This is consistent with the numerical work of Li \textit{et al} \cite{Li}  on the removal of CO$_2$, which exhibits a temperature rise of at most a few degrees, consequently we restrict the present  analysis to the isothermal case.

In keeping with standard practice we will work with cross-sectionally averaged quantities. Careful derivations of related averaged equations and their accuracy may be found in \cite{Aguareles23,Mondal19,Myers23}.

%%%%%%%%%%%%%%%%%%%%%%%%%%%%%%%%%%%%%%%%%%%%%%%%%%%%%%%%%%%%%%%%%%%%%%%%%%%%%%%%%%%%%%%%

\subsection{Governing equations}
Let ${c}_1,{c}_2$ be the concentration of contaminant and carrier gas. In the present analysis we work with mol/m$^3$, but kg/m$^3$ is also common in column adsorption studies, the conversion is trivial.  
Following \cite{Myers20}, the system of equations is
\begin{subequations}\label{DimEqs}
    \begin{align}
\pad{ c_1}{t}  + \pad{}{x}(u c_1  ) &= D  \padd{ c_1 }{x}  -  \frac{1-\epsilon}{\epsilon}\rho_a   \pad{q}{t}   ~ ,\label{c1eq} \\
\pad{c_2 }{t}  + \pad{}{x}(u c_2) &= D  \padd{ c_2}{x} \, \label{c2eq}.\\
p  & =    R_g T  ( c_1 +   c_2) \, ,\label{peq}\\
-\pad {p}{ x} &= \alpha (M_1 c_1 + M_2 c_2) u^2 + \left(\beta +\frac{(1-\epsilon)}{\epsilon}\rho_a M_1   \pad{q}{t}\right) u
 ~ , \label{pxeq}\\
 \pad{q}{t}&=\dot Q ~ ,\label{qeq}
\end{align}
\end{subequations}
where $D$ is the diffusion coefficient, $\rho_a$ is the particle apparent density, $R_g=8.31$ J/K$\cdot$mol is the universal gas constant, $T$ is the temperature, $M_1,M_2$ are the molar masses of the contaminant and the carrier gas respectively. The void fraction $\epsilon$ is assumed constant throughout the column.  The constants $\alpha$ and $\beta$ stem from the Ergun relation \cite{Myers20}, and are defined by
\begin{equation}
    \alpha = \frac{1.75(1-\epsilon)}{d_p\epsilon}\, ,\qquad \beta= \frac{150\mu_g(1-\epsilon)^2}{d_p^2\epsilon^2}\, ,
\end{equation}
where $d_p$ is the diameter of the adsorbing particles and $\mu_g$ is the gas viscosity. 

Note, the original paper by Myers \textit{et al.} \cite{Myers20} used slightly different notation and fitted for the unknown adsorbed material density. Here we will fit to the more standard adsorption coefficient. 

%the dividing $\epsilon$ in the sink term is missing, although it appears in their previous work \cite{Myers19} since it appears after averaging the mass conservation equations. Thus, this factor is taken into account in Eqs.~\eqref{c1eq} and \eqref{pxeq} for consistency with the units of the balances.}

%\textcolor{blue}{Previous papers by Myers \textit{et al.} \cite{Myers19,Myers20} also account for the density of the adsorbed contaminant ($\rho_q$). They determine its value in Myers \textit{et al.} \cite{Myers19} in order to fit the breakthrough curve. They state that this is the only fitting parameter since the adsorption coefficient is obtained using two points of the central part of the curve. Actually, this is a first fitting of the adsorption coefficient which they later correct by a second fitting of $\rho_q$. The correct density used should be the particle apparent density $\rho_a$ instead, which is an experimental value, and the adsorption coefficient should be fitted then to agree with the breakthrough data. The particle apparent density $\rho_a$ is only used to convert the mass density units of the adsorbed fraction into volumetric density. Note that the adsorbed fraction is defined over the total initial adsorbent mass inside the column, that remains invariable. Thus, it can be readily converted to mass per cubic meter of bed by multiplying by the bulk density $\rho_b$, which is the total mass of adsorbent divided by the column volume. This is related to the particle apparent density as $\rho_b=(1-\epsilon)\rho_a$.}

\subsection{Boundary and initial conditions}

The inlet condition is not trivial: we assume that the pressure at the inlet is higher than the pressure at the outlet, i.e., 
\begin{equation}
    p_{in}=p_a+\Delta p\,  ,
\end{equation}
where $p_a$ is the ambient (outlet) pressure and $\Delta p$ is the pressure drop, which must be a function of time to ensure that the gas velocity at the inlet remains constant. Using the ideal gas law, for a given volume fraction $\phi_i$, the concentration of species $i$ just upstream of the inlet is
\begin{align}\begin{split}\label{InletConcentrations}
    c_{i}(0^-,t)=&\frac{\phi_ip_{in}(t)}{R_gT}=\frac{\phi_i}{R_gT}(p_a+\Delta p)
    =\frac{\phi_ip_a}{R_gT}\left(1+\frac{\Delta p}{p_a}\right)=c_{i0}\left(1+\frac{\Delta p}{p_a}\right)\, ,
\end{split}\end{align}
where $T$ is the temperature and $R_g$  the universal gas constant. If the pressure drop is small ($\Delta p\ll p_a$) then the inlet concentrations may be considered approximately constant, i.e., $c_i(0^-,t)\approx c_{i0}$. We will restrict ourselves to this case since most practical applications have very low velocities and therefore very small pressure drops along the column.

The initial and boundary conditions are then: 
\begin{align}\label{bcConcentrations}
    u_0c_{i}(0^{-},t)    =   \left. \left(uc_i - D \pad{c_i}{x}\right)\right|_{x=0^+} \, ,\qquad \left.\pad{c_i}{x}\right|_{x=L}=0 \, ,
\end{align}
\begin{align}
\label{bcPressure}
     p(0,t) = p_a + \Delta p\, ,\qquad p(L,t)=p_a\, ,
\end{align}
with $i=1,2$. 
\\
Initially, there is neither free nor adsorbed contaminant in the column, hence
\begin{subequations}\label{initialconds}
    \begin{equation}
    c_1(x,0)=q(x,0)=0\, .
\end{equation}
The available space in the column is initially occupied by the carrier gas. From Eq. \eqref{peq} we find
\begin{equation}
    c_2(x,0)=\frac{p(x,0)}{R_gT}\, ,
\end{equation}
while for the pressure we impose an initially linear profile
\begin{equation}
    p(x,0)=p_a+\left(1-\frac{x}{L}\right)\Delta p(0)\, .
\end{equation}
\end{subequations}

\subsection{Choice of kinetic equation}

The function $\dot Q$ is key to correctly capturing the adsorption behaviour. Following the work of \cite{Dantas,Dantas2,Li} in \cite{Myers20}, the authors use the linear driving force  model 
\begin{equation}\label{LinearSink}
    \dot Q(q)=k_L(q_e-q)\, ,
\end{equation}
where $k_L$ is the adsorption rate for this particular model. As previously mentioned, this simple model has physical issues. For instance adsorption occurs whenever $q<q_e$, regardless of  whether contaminant is present or not. 

Here we employ a more general form, which may describe both chemical and physical adsorption, namely the Sips equation \cite{sips1948}
\begin{equation}\label{SipsSink}
    \dot Q(q,c_1)=k_ac_1^m(q_{max}-q)^n-k_dq^n\, ,
\end{equation}
where $k_a$ 
and $k_d$ 
are the adsorption and desorption rates respectively and we assume only component 1 is adsorbed. The maximum amount that may be adsorbed is denoted $q_{max}$, in general this is greater than the amount actually adsorbed. The powers $m$ and $n$ are related to the partial orders of reaction of the contaminant and the available adsorption sites in the global reaction
\begin{equation}
    m\mathbb{C}+n\mathbb{A} \rightleftharpoons \mathbb{AC}\, ,
\end{equation}
with $\mathbb{C}$ and $\mathbb{A}$ referring to the free contaminant and adsorbent molecules and $\mathbb{AC}$ to the reaction output. We will assume $m, n$ to be integers but note that with a chain of reactions it is possible that they take non-integer values.
% Note, for in the case of multiple reactions which may occur simultaneously or successively, $m$ and $n$ don't need to be integers necessarily CITE SOMETHING HERE.
The well-known Langmuir sink, describing physisorption, may be retrieved in the case $(m,n)=(1,1)$. 

When adsorption balances desorption, $\dot Q = 0$, the isotherm  is obtained. Denoting the equilibrium adsorbed amount $q_e$ for a given inlet contaminant concentration $c_{10}$, equation \eqref{SipsSink} may be rearranged to define the Sips isotherm
\bea
q_e=\frac{q_{max}K^{1/n}c_{10}^{m/n}}{1+K^{1/n}c_{10}^{m/n}} \, ,
\label{Sipsiso}
\eea
where $K=k_a/k_d$ is the equilibrium constant, whose units (m$^{3m}$kg$^{n-1}$mol$^{1-m-n}$) depend on the value of $m$ and $n$. The values of $q_{max}$ and $K$ may be determined experimentally from the recorded isotherms, as will be shown in \S\ref{sec:AnalyticsVSExperiments}. %This also allows determining which combination of $m,n$ will work best, although this may be inferred by observing the dominant chemical reaction. 

%%%%%%%%%%%%%%%%%%%%%%%%%%%%%%%%%%%%%%%%%%%%%%%%%%%%%%%%%%%%%%%%%%%%%%%%%%%%%%%%%%%%%%%%

%%%%%%%%%%%%%%%%%%%%%%%%%%%%%%%%%%%%%%%%%%%%%%%%%%%%%%%%%%%%%%%%%%%%%%%%%%%%%%%%%%%%%%%%

\section{Model Simplification}\label{sec:simplification}

At present the model consists of Eqs. \eqref{DimEqs}, with $\dot Q$  defined in Eq. \eqref{SipsSink}. The boundary and initial conditions are defined in Eqs. (\ref{bcConcentrations}--\ref{initialconds}). In this state it is difficult to make any analytical progress, consequently we now examine the relative size of terms by non-dimensionalising the system.

%%%%%%%%%%%%%%%%%%%%%%%%%%%%%%%%%%%%%%%%%%%%%%%%%%%%%%%%%%%%%%%%%%%%%%%%%%%%%%%%%%%%%%%%

\subsection{Non-dimensional formulation}
\label{subsection:NDform}
We introduce the scaled variables
\begin{equation}
    \hat{c}_1 = \frac{c_1}{c_{10}}\, ,\quad \hat{c}_2 = \frac{c_2}{c_{20}}\, ,\quad \hat{q} = \frac{q}{q_{max}}\, ,\quad \hat{u} = \frac{u}{\mathcal{U}}\, ,\quad \hat{x} = \frac{x}{\mathcal{L}}\, ,\quad \hat{p} = \frac{p-p_a}{\mathcal{P}}\, ,\quad \hat{t} = \frac{t}{{\mathcal{T}}}\, ,
\end{equation}
where the scales $\mathcal{U}$, $\mathcal{L}$, $\mathcal{P}$ and $\mathcal{T}$ are yet to be determined. The velocity scale is chosen as the constant inlet value $\mathcal{U}=u_0$. Since the focus is on the adsorption process, we choose $\mathcal{T}=1/(k_ac_{10}^mq_{max}^{n-1})$ to balance the time derivative and  adsorption terms in Eq. \eqref{SipsSink}. The length scale is determined by balancing the advection and sink terms in  Eq. \eqref{c1eq},  $\mathcal{L}=\epsilon\mathcal{U}\mathcal{T} c_{10}/[(1-\epsilon)\rho_a q_{max}]$. 
The pressure scale can be determined through Eq. \eqref{pxeq}. Upon replacing the dimensional quantities by their scaled counterparts, the Ergun equation \eqref{pxeq} becomes
\begin{align}\label{ndpx}
-\mathcal{P}\pad {\hat p}{ \hat x} &= \alpha\mathcal{L} \mathcal{U}^2(M_1 c_{10}\hat c_1 + M_2 c_{20}\hat c_2) \hat u^2 + \beta\mathcal{L} \mathcal{U}\hat u+\frac{(1-\epsilon)\rho_a M_1q_{max}\mathcal{L}\mathcal{U}}{\epsilon\mathcal{T}}\hat u\pad{\hat q}{\hat t} \, .
\end{align}
Assuming approximately Darcy flow, we choose $\mathcal{P}=\beta\mathcal{L}\mathcal{U}$ to balance the pressure gradient and the second term on the right-hand side. 

The final set of non-dimensional equations is now
\begin{subequations}\label{nondimEqs}
\begin{align}
    \delta_1\pad{\hat{c}_1}{\hat t}+\pad{}{\hat x}\left(\hat u\hat c_1\right)&=\delta_2\padd{\hat c_1}{\hat x}-\pad{\hat q}{\hat t}\, ,\label{ndc1eq}\\
    \delta_1\pad{\hat{c}_2}{\hat t}+\pad{}{\hat x}\left(\hat u\hat c_2\right)&=\delta_2\padd{\hat c_2}{\hat x}\, ,\label{ndc2eq}\\
    1+\delta_3\hat p  & =    \phi_1\hat c_1 +\phi_2\hat c_2\, ,\label{ndpeq}\\
    -\pad {\hat p}{ \hat x} &= \delta_4 \left(\delta_5\hat c_1 + \hat c_2\right) \hat u^2 + \left(1 +\delta_6\pad{\hat q}{\hat t}\right) \hat u\, ,\label{ndpxeq}\\
    \pad{\hat q}{\hat t}&=\hat c_1^m(1-\hat q)^n-\delta_7\hat q^n\, .\label{ndqeq}
\end{align}
\end{subequations}
These are subject to the boundary and initial conditions
\begin{subequations}\label{nondimBCs}
\begin{align}
    &1+\delta_3 \hat p    =   \left. \left(\hat u\hat c_1 - \delta_2 \pad{\hat c_1}{\hat x}\right)\right|_{\hat x=0} \, ,\qquad \left.\pad{\hat c_1}{\hat x}\right|_{\hat x=\hat L}=0\, ,\qquad \hat c_1(\hat x,0)=0\, ,\label{nondimBCs1}\\
    &1+\delta_3 \hat p    =  \left. \left( \hat u\hat c_2 - \delta_2 \pad{\hat c_2}{\hat x}\right)\right|_{\hat x=0}  \, ,\qquad \left.\pad{\hat c_2}{\hat x}\right|_{\hat x=\hat L}=0\, ,\qquad \phi_2 \hat c_2(\hat x,0)=1+\delta_3\hat p_{in}(\hat x)\, ,\label{nondimBCs2}\\
    &
    %\hat p(0,\hat t) = \Delta\hat p\, ,\qquad 
    \hat p(\hat L,\hat t)=0\, ,\qquad \hat p(\hat x,0)=\hat p_{in}(\hat x)=\Delta\hat p(0)\left(1-\frac{\hat x}{\hat L}\right)\, ,\label{nondimBCs3}\\  
    &\hat q(\hat x,0)=0\, ,\label{nondimBCs4}
\end{align}
\end{subequations}
with $\Delta\hat p(0)=\Delta p(0)/\mathcal{P}$. Evaluating  \eqref{ndpxeq} at $t=0$ and combining it with \eqref{nondimBCs3}, gives 
\begin{align}\label{nondimICu}
 \delta_4 \hat c_2(\hat{x},0)\, \hat{u}(\hat{x},0)^2 +  \hat{u}(\hat{x},0) - \frac{\Delta\hat p(0)}{\hat{L}} = 0\,. 
\end{align}
The positive roots of the polynomial \eqref{nondimICu} provide an initial profile for $\hat{u}$. 

Besides the volume fractions $\phi_1$ and $\phi_2$, the non-dimensional model has 7 additional dimensionless quantities, defined by
\begin{align}\begin{split}\label{nondimparams}
    &\delta_1=\frac{\mathcal{L}}{\mathcal{U}\mathcal{T}}\, ,\qquad \delta_2=\frac{D}{\mathcal{L}\mathcal{U}}\,\qquad \delta_3=\frac{\mathcal{P}}{p_a}\, ,\qquad 
    \delta_4=\frac{\alpha M_2c_{20}\mathcal{U}^2\mathcal{L}}{\mathcal{P}}\, ,\\ 
    \delta_5=&\frac{c_{10}M_1}{c_{20}M_2}\, ,\qquad \delta_6=\frac{(1-\epsilon)\rho_aM_1q_{max}\mathcal{L}\mathcal{U}}{\epsilon\mathcal{T}\mathcal{P}}\, ,\qquad \delta_7=\frac{k_d}{k_ac_{10}^m}=\frac{1}{Kc_{10}^m}\, .
\end{split}\end{align}
The value of $k_a$ is not easily directly determined by experiments, hence we consider it a free parameter in the current model. 
%This quantity is included in the definitions of $\delta_1,\delta_2,\delta_3,\delta_4$ and $\delta_6$ (through $\mathcal{T}$, $\mathcal{L}$ and $\mathcal{P}$). After some algebra, we can rewrite these as
Replacing the scales we obtain
\begin{align}\begin{split}\label{nondimparamsAlt}
    &\delta_1=\frac{\epsilon c_{10}}{(1-\epsilon)\rho_a q_{max}}\, ,\qquad 
    \delta_2=\left(\frac{(1-\epsilon)\rho_aDc^{m-1}_{10}q_{max}^n}{\epsilon u_0^2}\right)k_a\,\\ 
    &\delta_3=\left(\frac{\epsilon\beta u_0^2}{(1-\epsilon)\rho_ap_ac^{m-1}_{10}q_{max}^n}\right)\frac{1}{k_a}\, ,\\
    &\delta_4=\frac{\alpha M_2c_{20}u_0}{\beta}\, ,\qquad 
    \delta_6=\left(\frac{(1-\epsilon)\rho_aM_1c^m_{10}q_{max}^n}{\epsilon\beta}\right)k_a\, .
\end{split}\end{align}
The non-dimensional equilibrium value $\hat q_e=q_e/q_{max}$ is 
\bea\label{def:NDqe}
\hat q_e=\frac{1}{1+\delta_7^{1/n}} \, .
\label{NDiso}
\eea

%%%%%%%%%%%%%%%%%%%%%%%%%%%%%%%%%%%%%%%%%%%%%%%%%%%%%%%%%%%%%%%%%%%%%%%%%%%%%%%%%%%%%%%%

\subsection{Interpretation of parameters}\label{ssec:parameters}
Since $\mathcal{L}$ and $\mathcal{T}$ are both chosen to balance the adsorption terms, the ratio $\mathcal{L}/\mathcal{T}$ may be interpreted as an adsorption velocity. Therefore, the parameter $\delta_1=\mathcal{L}/(\mathcal{T} u_0)$ is the ratio of the adsorption velocity to the fluid velocity and hence  represents a form of Damk\"ohler number. 
The non-dimensional parameter $\delta_2=D/(\mathcal{L}u_0)$ represents the ratio of the diffusive transport rate $D/\mathcal{L}$ to the advective transport rate $u_0$ and is therefore an inverse P\`eclet number. 
The third parameter $\delta_3=\mathcal{P}/p_a$ is the Darcy pressure scale relative to the ambient pressure. 
In \S\ref{sec:AnalyticsVSExperiments} we demonstrate that for the cases studied, which are representative of adsorption from a gas stream,  $\delta_1=\ord{10^{-3}}$, $\delta_2=\ord{10^{-1}}$ or smaller and $\delta_3$ is  $\ord{10^{-2}}$ or smaller.

The parameter $\delta_4$ relates the kinetic energy of the carrier fluid to the pressure gradient driving the flow or viscous resistance (equivalent since the two are balanced). Then  $\delta_6$ relates the energy loss due to mass removal to the viscous resistance. Since the flow is driven by the pressure gradient we expect $\delta_4,\delta_6\ll1$. In \S\ref{sec:AnalyticsVSExperiments} they are shown to be  $\ord{10^{-3}}$.

The parameter $\delta_5$ represents the ratio of densities and will vary depending on the contaminant, carrier fluid and volume fractions. We may assume this quantity to be $\ord{1}$.  

Finally, $\delta_7=k_d/(k_a c_{10}^m)$ represents the ratio of the adsorption to the desorption time scales. If the value of $\delta_7$ is small, it suggests that adsorption dominates. Although our focus is on adsorption processes we will retain it in the reduced system of equations to permit modelling of  filters where both processes are comparable or to include regeneration of filters. If its value is significantly very small then we lose nothing through its retention. It is also important to retain $\delta_7$ to correctly model the isotherm and determine the value $k_d/k_a$.

\section{Solution methods}\label{sec:SolutionMethods}
In this section we first take advantage of the anticipated small size of certain terms to reduce the problem to a more tractable set of equations (the size of neglected terms is verified in \S \ref{sec:results}). By applying a travelling wave substitution we are then able to obtain exact solutions for the  cases where $m=1$ and $n=1,2$. Further solutions may be possible, as is the case in the study of the removal of trace elements described in \cite{Aguareles23}, however our experimental data only covers these two cases. Subsequently we briefly describe a numerical scheme for a more complete set of equations.

\subsection{Approximate solution method}\label{sec:AnalyticalSolns}

Neglecting the $\ord{\delta_3}$ term in  Eq. \eqref{ndpeq} and applying the relation $\phi_2=1-\phi_1$, we obtain a simple relation between the concentrations
\begin{equation}\label{redc2}
    \hat c_2 = \frac{1-\phi_1\hat c_1}{1-\phi_1}\, .
\end{equation}
If $\delta_3 \sim 10^{-2}$ we expect this reduction to result in errors of the order 1\%.

Upon neglecting terms of order $\ord{\delta_1,\delta_2}$, the mass balance for the carrier fluid becomes
\begin{equation}
    \pad{}{\hat x}\left(\hat u\hat c_2\right)=0\, ,\label{ndc2eqred}
\end{equation}
subject to $\hat u\hat c_2=1$ at $\hat x=0$ (this is the reduced form of condition \eqref{nondimBCs2} after also neglecting $\delta_3$). This trivially leads to 
\begin{equation}\label{redu}
    \hat u\hat c_2=1\Rightarrow \hat u = \frac{1}{\hat c_2}=\frac{1-\phi_1}{1-\phi_1\hat c_1}\, .
\end{equation}
% where $\phi_{1}=\delta_4\delta_5$. 
After combining Eqs. \eqref{ndc1eq}, \eqref{redu} and \eqref{redc2} and neglecting terms of order $\ord{\delta_1,\delta_2}$, we obtain
\begin{equation}
    \pad{}{\hat x}\left(\frac{(1-\phi_1)\hat c_1}{1-\phi_1\hat c_1}\right)=-\pad{\hat q}{\hat t}\, ,\label{ndc1eqred}
\end{equation}
which must be solved along with
\begin{equation}
    \pad{\hat q}{\hat t}=\hat c_1^m(1-\hat q)^n-\delta_7\hat q^n\, .\label{ndqeqred}
\end{equation}
The pressure is determined by integrating 
\begin{equation}\label{ndpxeqred}
    -\pad {\hat p}{ \hat x} = \hat u\, ,
\end{equation}
(neglecting terms of order $\ord{\delta_4,\delta_6}$). Using $\hat p(\hat L,\hat t)=0$ we find
\begin{equation}
    \hat p(\hat x, \hat t) = \int_{\hat x}^{\hat L}\hat u(\hat \xi,\hat t)\ud \hat \xi\, ,
\end{equation}
therefore the system has now been reduced to solving for only two unknowns, $\hat c_1$ and $\hat q$. 

The reduced system, Eqs.~(\ref{ndc1eqred},\ref{ndqeqred}), may be solved exactly for certain combinations of $m$ and $n$.
This involves introducing a function $F (\eta) = (1-\phi_1) \hat c_1/(1-\phi_1 \hat c_1)$ where the travelling wave co-ordinate $\eta = \hat x - \hat s(\hat t)$, and $d \hat s/d \hat t = v$ is constant,
see  \cite{Myers20,Myers23,Aguareles23}.  In the travelling wave co-ordinate system the equations may be integrated immediately. The derivation for the most common physical cases $m=1$ and $n=1,2$ is detailed in the Supplementary Information (these values correspond to the experiments discussed in \S\ref{sec:AnalyticsVSExperiments}). The solution for the contaminant concentration $\hat c_1(\hat x, \hat t)$ may be written in implicit form
\begin{equation}
\label{ndBreakthroughModel}
    \left(1+\delta_7^{1/n}\right)\left(\hat t-\hat t_{1/2}\right) - \left(\hat x-\hat L\right) = \hat Y_{mn}\left(\hat c_1\right)\, ,
\end{equation}
where
% In the case $(m,n)=(1,1)$, we find
% \begin{subequations}\label{TWsolution:Y}
% \begin{align}
%     -\hat\eta &= \frac{1+\delta_7}{1+\phi_1\delta_7}\left[(1-\phi_1)\ln\left(\frac{\hat F}{\hat F_{1/2}}\right)
%     -\ln\left(\frac{1-\hat F}{1-\hat F_{1/2}}\right)\right]\, ,
% \end{align}
% whereas for $m=1$ and $n=2$, the solution to Eq. \eqref{TW_final} is given by
% \begin{align}
%     -\hat\eta &= (1-\phi_1)\ln\left(\frac{\hat F}{\hat F_{1/2}}\right) - \frac{a}{a-1}\ln\left(\frac{1-\hat F}{1-\hat F_{1/2}}\right) + \left(\phi_1+\frac{1}{a-1}\right)\ln\left(\frac{a-\hat F}{a-\hat F_{1/2}}\right)\, ,
% \end{align}
% \end{subequations}
% where $a=(1+\sqrt{\delta_7})^2/(1-\phi_1\delta_7)$.
% Upon reverting the definition of $\hat F$, we can write implicit solutions of the form $-\hat\eta=\hat Y_n^m(\hat c_1)$, with
\begin{subequations}\label{TWsolution:Yc}
\begin{align}
    \hat Y_{11}(\hat c_1) &= \frac{1+\delta_7}{1+\phi_1\delta_7}\left(\ln\left|\frac{\hat c_1}{1-\hat c_1}\right|-\phi_1\ln\left|\frac{(2-\phi_1)\hat c_1}{1-\phi_1\hat c_1}\right|\right)\, ,\\
    \hat Y_{12}(\hat c_1) &= \frac{1+\sqrt{\delta_7}}{\sqrt{\delta_7}\left(1+\sqrt{\gamma}\right)}\left(\ln\left|\frac{\hat c_1}{1-\hat c_1}\right|-\gamma\ln\left|\frac{(2-\gamma)\hat c_1}{1-\gamma\hat c_1}\right|\right)\, ,
\end{align}
and
\begin{equation}
\gamma=\left(\frac{1+\phi_{1}\sqrt{\delta_7}}{1+\sqrt{\delta_7}}\right)^2 \, . \label{eq:gamma}
\end{equation}\end{subequations}
The remaining variables may be calculated from the relations
\begin{align}\label{TW:remaining}
    \hat c_2&=\frac{1-\phi_1\hat c_1}{1-\phi_1}\, ,\quad 
    \hat q = \frac{(1-\phi_1) \hat c_1}{(1+\delta_7^{1/n})(1-\phi_1 \hat c_1)} 
    %=  \frac{\hat u \hat c_1}{(1+\delta_7^{1/n})}
    \, ,\quad
%    \hat q = \frac{\hat c_1}{1+\delta_7^{1/n}}
%    \, ,\qquad
    \hat u = \frac{1-\phi_1}{1-\phi_1\hat c_1}\, ,\quad
     \hat p = \int_{\hat x}^{\hat L}\hat u(\hat \xi,\hat t)\ud \hat \xi\, 
    .
\end{align}
Setting $\hat x=\hat L$ in \eqref{ndBreakthroughModel} provides the  breakthrough curve
\begin{equation}\label{BreakthroughModel}
    \hat t = \hat t_{1/2} + \frac{\hat Y_{mn}\left(\hat c_{1b}\right)}{1+\delta_7^{1/n}} \, ,
\end{equation}
with $\hat c_{1b}(\hat t)=\hat c_1(\hat L,\hat t)$.
The fact that this solution depends only on $\delta_1$ and $\phi_1$,  indicates that the most important parameters affecting  breakthrough  are the ratio of adsorption to desorption, the feed contaminant concentration and volume fraction.

\subsection{Dimensional solutions}
The solutions given by equations (\ref{ndBreakthroughModel}--\ref{BreakthroughModel}) define the concentration of contaminant, carrier fluid, amount adsorbed, velocity and pressure. All of these solutions are novel and represent a breakthrough in the understanding of the adsorption of large quantities of a fluid. Given that many practitioners prefer to work in dimensional form we now present the solutions in terms of the original variables.

The concentration of the contaminant is 
\begin{equation}\label{Soln:c1}
    \left(1+\delta_7^{1/n}\right)\frac{t-t_{1/2}}{\mathcal{T}} - \frac{x-L}{\mathcal{L}} =   Y_{mn}\left(c_1\right)\, ,
\end{equation}
where
\begin{subequations}
\begin{align}
      \label{Y11Dim}
      Y_{11}(  c_1) &= \frac{1+\delta_7}{1+\phi_1\delta_7}\left(\ln\left|\frac{  c_1}{c_{10}-  c_1}\right|-\phi_1\ln\left|\frac{(2-\phi)  c_1}{c_{10} -\phi_1  c_1}\right|\right)\, ,\\
      \label{Y12Dim}
     Y_{12}(c_1) &= \frac{1+\sqrt{\delta_7}}{\sqrt{\delta_7}\left(1+\sqrt{\gamma}\right)}\left(\ln\left|\frac{  c_1}{c_{10}- c_1}\right|-\gamma\ln\left|\frac{(2-\gamma)  c_1}{c_{10} -\gamma  c_1}\right|\right)\, ,
\end{align}\end{subequations}
and $\gamma$ defined by equation \eqref{eq:gamma}.

The carrier gas concentration, adsorbed amount, and velocity are then 
\begin{subequations}\begin{align}\label{Soln:others}
    {c_2}(x,t)&=\frac{p_a}{R_gT}-c_1(x,t)\, ,\, \, \,  
    %,\\
       q(x,t) = \frac{q_{max}(1-\phi_1)  c_1(x,t)}{(1+\delta_7^{1/n})(c_{10}-\phi_1 \hat c_1)}  \, ,
   % q(x,t) = \frac{q_{max}\, c_1(x,t)}{c_{10}\left(1+\delta_7^{1/n}\right)}\, ,
    %,\\
     \, \, \, u(x,t) = \frac{(1-\phi_1)u_0c_{10}}{c_{10}-\phi_{1}c_1(x,t)}\, ,
\end{align}\end{subequations}
while the pressure is given by
\begin{equation}\label{Soln:p}
    p(x,t) = p_a+\beta\int_{x}^{L} u(\xi,t)\ud \xi\, .
\end{equation}

Finally, evaluating Eq. \eqref{Soln:c1} at $x=L$ yields the dimensional breakthrough models for the Langmuir model, $(m,n)=(1,1)$
\begin{subequations}
\label{BC_all}
\begin{equation}
\label{BC11}
    t=t_{1/2}+\frac{\mathcal{T}}{1+\phi_{1}\delta_7}\left(\ln\left|\frac{c_{1b}}{c_{10}-c_{1b}} \right|-\phi_{1}\ln\left|\frac{\left(2-\phi_{1}\right)c_{1b}}{c_{10}-\phi_{1}c_{1b}} \right|\right) \, ,
\end{equation}
and the Sips model with $(m,n)=(1,2)$ 
\begin{equation}
\label{BC12}
    t=t_{1/2}+\frac{\mathcal{T}}{\sqrt{\delta_7}\left(1+\sqrt{\gamma}\right)}\left(\ln\left|\frac{c_{1b}}{c_{10}-c_{1b}} \right|-\gamma\ln\left|\frac{\left(2-\gamma\right)c_{1b}}{c_{10}-\gamma c_{1b}} \right|\right) \, ,
\end{equation}\end{subequations}
where $c_{1b}=c_1(L,t)$ is the breakthrough concentration of component 1, $\mathcal{T}=1/\left(k_ac_{10}q_{max}^{n-1}\right)$, $\delta_7=1/(Kc_{10})$ and $\gamma=\left((1+\phi_1\sqrt{\delta_7})/(1+\sqrt{\delta_7})\right)^2$.

In the limit of trace amounts of contaminant, $\phi_{1}\rightarrow 0$, equation \eqref{BC11}  reduces to 
\begin{equation}
\label{LangSol}
  t=t_{1/2}+\frac{1}{k_ac_{10}} \ln\left|\frac{c_{1b}}{c_{10}-c_{1b}} \right|\ \,  \quad \Ra \quad c_{1b} = \frac{c_{10}}{1+ \exp(k_ac_{10}(t_{1/2}-t))} \, ,
\end{equation}
which is the result of Myers \emph{et al.} \cite{Myers23}. 
Setting $\phi_{1}\rightarrow 0$, equation \eqref{BC12} with $\mathcal{T}=1/\left(k_ac_{10}q_{max}\right)$, $\gamma=(1+\sqrt{\delta_7})^{-2}$ reproduces the $(m,n)=(1,2)$ result of Aguareles \emph{et al.} \cite{Aguareles23}.

%%%%%%%%%%%%%%%%%%%%%%%%%%%%%%%%%%%%%%%%%%%%%%%%%%%%%%%%%%%%%%%%%%%%%%%%%%%%%%%%%%%%%%%%

\subsection{Numerical solution}\label{sec:numerics}

Due to the non-linear form of Eq. \eqref{ndpxeq} the numerical integration of the full model is not straightforward. However, following the work of \cite{Myers20} we may develop a numerical scheme  neglecting only the  parameters $\delta_3,\delta_4,\delta_6$, which have a maximum size of order $10^{-2}$ (so we retain any larger parameters).   Eq. \eqref{ndpxeq} then becomes Eq. \eqref{ndpxeqred} and Eq. \eqref{ndpeq} can be rearranged to Eq. \eqref{redc2}. Substituting the expression for $\hat c_2$ into Eq. \eqref{ndc2eq} and combining with   Eq. \eqref{ndc1eq} 
yields
\begin{equation}\label{numEqu}
    \pad{\hat u}{\hat x}=\phi_1\left(\delta_{1}\pad{\hat{c}_1}{\hat t} +\pad{}{\hat x}\left(\hat u\hat c_1\right)-\delta_{2}\padd{\hat c_1}{\hat x}\right)=-\phi_1\pad{\hat q}{\hat t}\, .
\end{equation}
A boundary condition for $\hat u$ is obtained by combining conditions (\ref{nondimBCs2},  \ref{nondimBCs1}) at $\hat x=0$ along with the relation \eqref{redc2}
 to determine $\hat u(0,\hat t)=1$.  Integrating Eq. \eqref{numEqu} then gives
\begin{equation}\label{numSolnu}
    \hat u(\hat x,\hat t) = 1 - \phi_1\int_0^{\hat x}\pad{\hat q}{\hat t}(\hat \xi,\hat t)\ud \hat\xi=1-\phi_1\hat{Q}_{acc}(\hat x,\hat t)\, ,
\end{equation}
where $\hat{Q}_{acc}(\hat x,\hat t)$ represents the rate of change of accumulated contaminant within the column. Substituting this  into Eq. \eqref{ndpxeqred} and integrating yields
\begin{equation}\label{numSolnp}
    \hat p(\hat x,\hat t) = \hat L-\hat x -\phi_1\int_{\hat x}^{\hat L}\hat{Q}_{acc}(\hat \xi,\hat t)\ud\hat\xi\, ,
\end{equation}
where we have used $\hat p=0$ at $\hat x=\hat L$. 

Within this approximation the problem reduces to solving Eqs. \eqref{ndc1eq} and \eqref{ndqeq}, along with the integral \eqref{numSolnu}, to find $\hat{c}_1$, $\hat{q}$ and $\hat{u}$, while $\hat{c}_2$ and $\hat{p}$ become passive variables that may be obtained afterwards. For the numerical integration of \eqref{ndc1eq} and \eqref{ndqeq} we use an explicit Euler marching scheme, while the spatial derivatives from the advection and diffusion terms in \eqref{ndc1eq} are discretised via a first order upwind scheme and a second order central differences scheme, respectively.

%%%%%%%%%%%%%%%%%%%%%%%%%%%%%%%%%%%%%%%%%%%%%%%%%%%%%%%%%%%%%%%%%%%%%%%%%%%%%%%%%%%%%%%%

\subsection{Discussion of solutions}\label{sec:results}

In this section we show and discuss the analytical solutions obtained in section \ref{sec:AnalyticalSolns} and compare them with the numerical solutions using the approach described in section \ref{sec:numerics}. We also discuss the differences between the solutions of the current model and the corresponding ones when the flow velocity is  constant. 

%%%%%%%%%%%%%%%%%%%%%%%%%%%%%%%%%%%%%%%%%%%%%%%%%%%%%%%%%%%%%%%%%%%%%%%%%%%%%%%%%%%%%%%%

\subsubsection{Travelling wave solutions}

%To comment: (1) General description of variables (c1,c2,1,u) behaviour and comment differences between m=n=1 and n=2 and m=1. (2) Comparison to linear driving force. 

\begin{figure}
    \centering
    \begin{subfigure}[b]{0.48\textwidth}
         \centering
         \includegraphics[width=\textwidth]{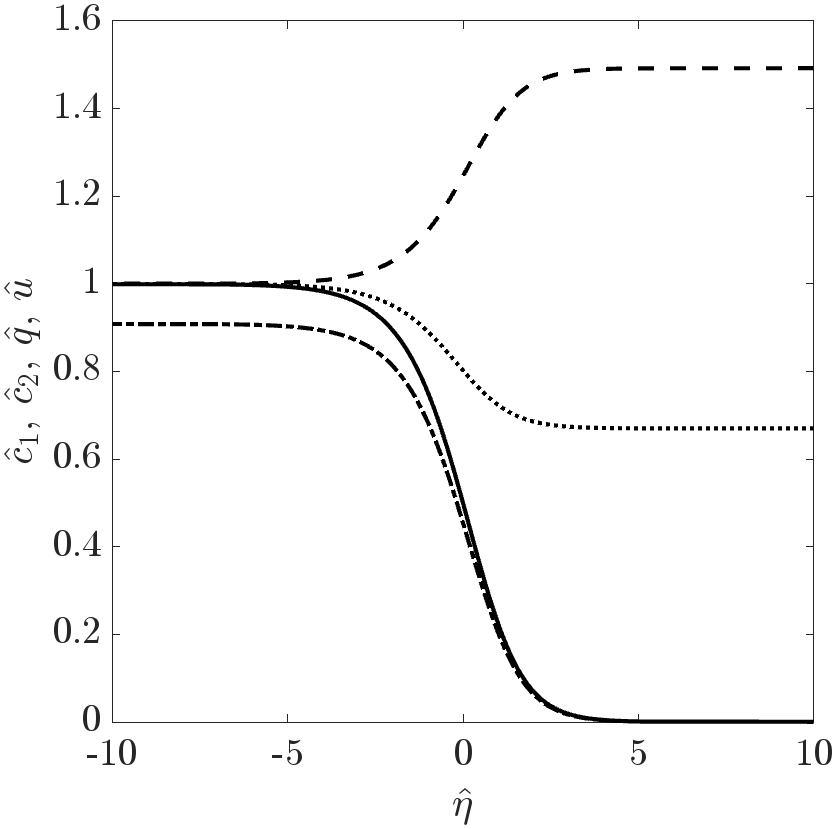}
         \caption{}
         \label{fig:TWsolutions_eta11_bw}
     \end{subfigure}
     \hfill
     \begin{subfigure}[b]{0.48\textwidth}
         \centering
         \includegraphics[width=\textwidth]{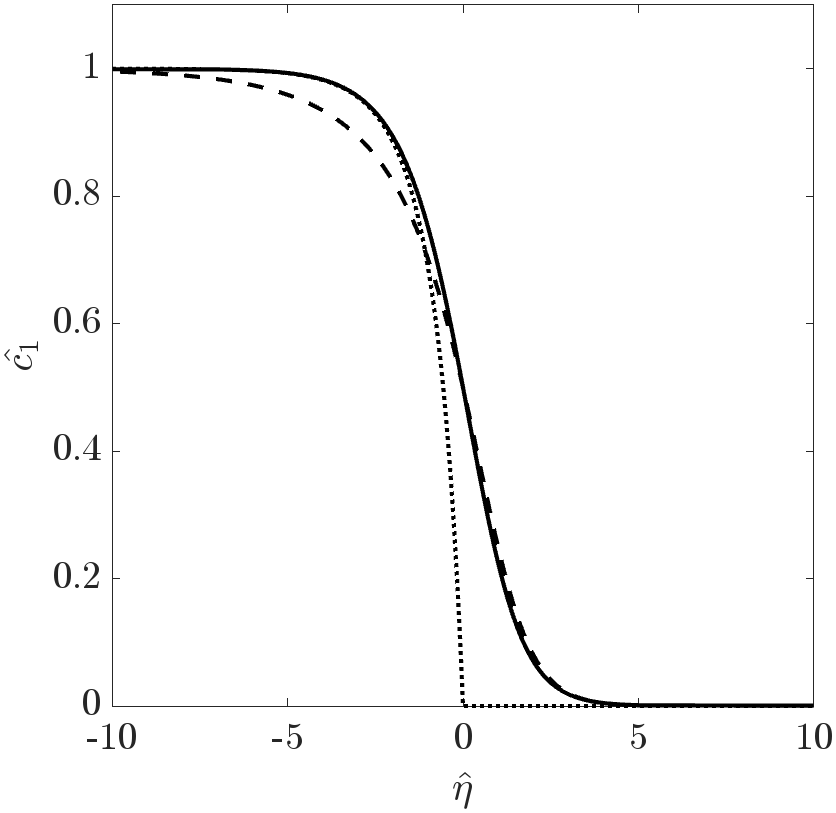}
         \caption{}
         \label{fig:TWsolutions:c1}
     \end{subfigure}
    % \includegraphics[width=.48\textwidth]{F_eta_bw.png}
    % % \includegraphics[width=.48\textwidth]{placeholder.png}
    % \hfill
    % \includegraphics[width=.48\textwidth]{c1_eta_bw.png}
    % % \includegraphics[width=.48\textwidth]{placeholder.png}
    \caption{Profiles of (a) the contaminant concentration $\hat c_1$ (solid line), the carrier fluid concentration $\hat c_2$ (dashed line), adsorbed amount $\hat q$ (dashed-dotted line) and flow velocity $\hat u$ (dotted line) for the case $(m,n)=(1,1)$ as a function of the travelling wave coordinate $\hat\eta$ with $\phi_1=0.33$ and $\delta_7=0.1$; (b) the contaminant concentration $\hat c_1$: solid and dashed lines correspond to the cases $(1,1)$ and $(1,2)$. For comparison, the linear driving force solution (LDF) presented in \cite{Myers20} is shown as  a dotted line.}
    \label{fig:TWsolutions:Fandc}
\end{figure}

In Figure \ref{fig:TWsolutions_eta11_bw} we show the profiles of $\hat{c}_1$, $\hat{c}_2$, $\hat{q}$ and $\hat{u}$ from the travelling wave solution as a function of the travelling wave coordinate $\eta$ for the case $(m,n)=(1,1)$. The concentration profile for the contaminant is characterised by a wave  that tends to the inlet value, $c_1=1$, as $\eta\rightarrow-\infty$ and decreases gradually to 0 as $\eta\rightarrow+\infty$. Similarly, the adsorbed fraction tends to its constant equilibrium value near the inlet, in this case $\hat q \approx 0.88$, and to zero near the wave front where the adsorbent material is still clean of contaminant (i.e, $\hat q = 0$). In the region of the column where the adsorbent is full of contaminant the gas velocity is highest and it decreases towards the clean region of the column, indicating that adsorption slows down the propagation of the gas mixture through the column. The profiles for the case $(1,2)$ (not shown) show equivalent behaviour with some qualitative differences in the concentration profiles.  
% (\textcolor{red}{mention qualitative differences, or refer to a Sup. Information document - or say we discuss differences for $c_1$ next}).

In Figure \ref{fig:TWsolutions:c1} we show the concentration profiles for the contaminant for the cases $(m,n)=(1,1)$ and $(1,2)$, along with the profile obtained by  the LDF model, equation \eqref{LinearSink}, which is derived in \cite{Myers20}. 
The profiles for $(m,n)=(1,1)$ and $(1,2)$ are almost identical for $\eta>0$ while for $\eta<0$ the case $(1,2)$ shows lower concentrations of contaminant than the case $(m,n)=(1,1)$. This is consistent with the results reported  in Aguareles et. al. \cite{Aguareles23} for the removal of trace amounts. The profile corresponding to the LDF model shows a very rapid decrease to zero at $\eta = 0$, this is related to its physical deficiency: near the wave front $c, q \ra 0$ and hence $\partial q/\partial t = k_L(q_e-q) \ra k_L q_e$, that is adsorption rate takes its highest possible value where there is virtually no contaminant.

%The concentration profile for the contaminant, $\hat{c}_1$, is characterised by a wave front shape that tends to 1 as $\eta\rightarrow-\infty$ and decreases gradually to 0 as $\eta\rightarrow+\infty$. This is consistent with the fact that the region of the adsorbent material near the inlet of the column has already been filled with contaminant and, therefore, the concentration of contaminant must match the inlet value. Conversely, the contaminant has not reached the end of the column, where the adsorbent remains clean, so $\hat{c}_1=0$. 

%The travelling wave front forms in a very short transient of order delta1, once formed it will travel from left to right throughout the column. 

%This is consistent with the fact that, after some very short transient,  

%The discussion for case $(m,n)=(1,2)$ is analogous. 

%%%%%%%%%%%%%%%%%%%%%%%%%%%%%%%%%%%%%%%%%%%%%%%%%%%%%%%%%%%%%%%%%%%%%%%%%%%%%%%%%%%%%%%%

\subsubsection{Breakthrough curves}

An accurate description of the breakthrough curve is crucial to the understanding of the mass transfer dynamics in column adsorption. The travelling wave solutions developed in section \S \ref{sec:AnalyticalSolns} present a simple method to describe the behaviour throughout the column and specifically the breakthrough. However, before applying the approximate solutions to real breakthrough data it is important to verify their accuracy. Here we do this through comparison with the numerical solution.

\begin{figure}
    \centering
    \begin{subfigure}[b]{0.48\textwidth}
         \centering
         \includegraphics[width=\textwidth]{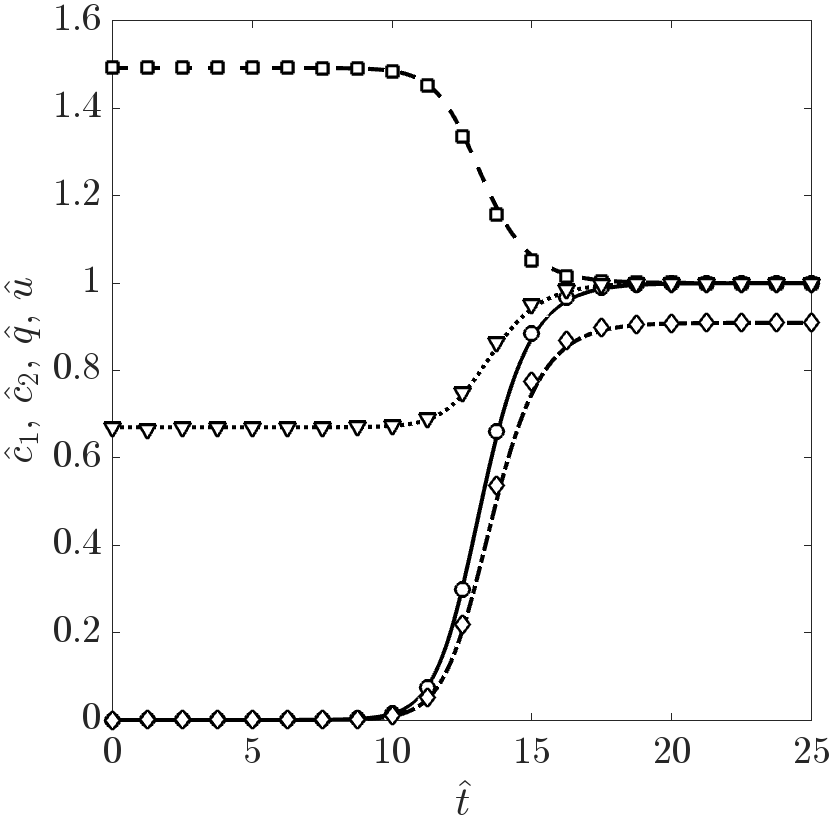}
         \caption{$(1,1)$}
         \label{fig:TWsolutions:all11}
     \end{subfigure}
     \hfill
     \begin{subfigure}[b]{0.48\textwidth}
         \centering
         \includegraphics[width=\textwidth]{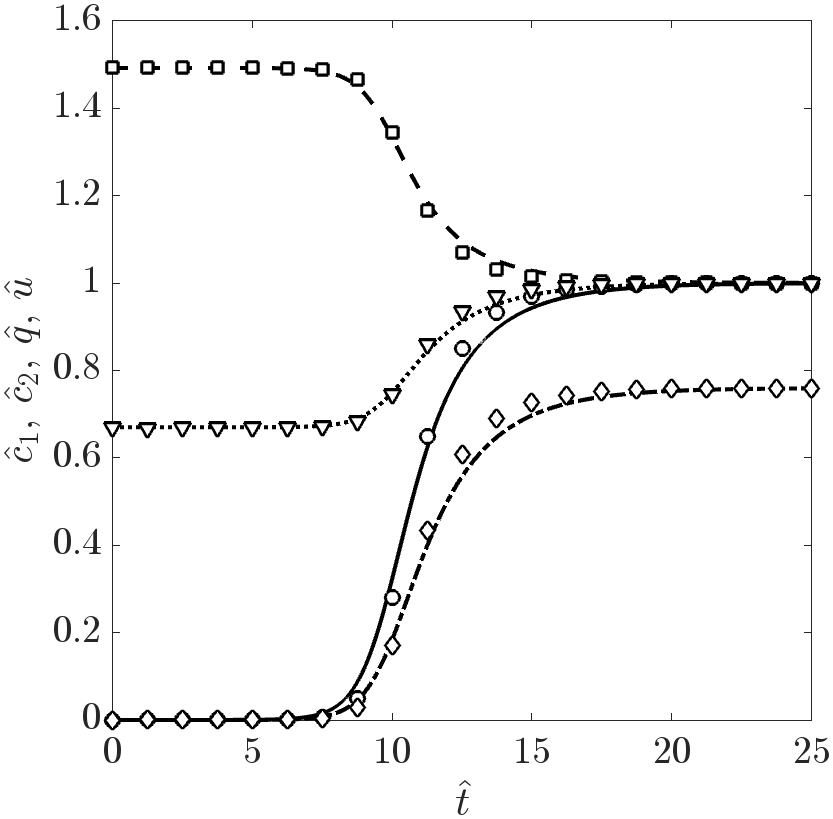}
         \caption{$(1,2)$}
         \label{fig:TWsolutions:all12}
     \end{subfigure}
    % \includegraphics[width=.48\textwidth]{All_eta11_bw.png}
    % % \includegraphics[width=.48\textwidth]{placeholder.png}
    % \hfill
    % \includegraphics[width=.48\textwidth]{All_eta12v2_bw.png}
    % % \includegraphics[width=.48\textwidth]{placeholder.png}
    \caption{Evolution of the  dependent variables at $\hat x=\hat L$ for different combinations $m=1$, $n=1, 2$. Lines refer to the analytical solutions developed in \S \ref{sec:AnalyticalSolns},  symbols refer to the numerical solutions described in \S \ref{sec:numerics}. The contaminant and carrier concentrations $\hat c_1$ and $\hat c_2$, adsorbed amount $\hat q$ and flow velocity $\hat u$ are represented, respectively, by solid lines and circles, dashed lines and squares, dashed-dotted lines and diamonds, and dotted lines and triangles. The parameters have been set to $\phi_1=0.33$ and $\delta_7=0.1$.  The remaining parameters in the numerical solutions are $\hat{L}=15$, $\delta_1=10^{-3}$, $\delta_2=0.01$, $\delta_3=0.01$, $\delta_4=10^{-3}$, $\delta_5=0.1$, and $\delta_6=10^{-5}$, which are consistent with the experimental data for CO$_2$ capture of \cite{Monazam2013}, see \S\ref{sec:AnalyticsVSExperiments}.}
    \label{fig:TWsolutions:all}
\end{figure}

In Figure \ref{fig:TWsolutions:all} we compare the analytical expressions for breakthrough of  $\hat{c}_1$, $\hat{c}_2$, $\hat{q}$ and $\hat{u}$  with those obtained numerically using the procedure described in \S \ref{sec:numerics}. In the case $n=1$ the results are barely distinguishable, for the case $n=2$ slight differences may be observed close to the end of the wave. We note that first breakthrough occurs at a slightly later time for the $n=1$ case and the curve $\hat c_1(\hat L, \hat t)$ is slightly steeper. This suggests a lower (non-dimensional) average velocity of the carrier fluid, which is consistent with the contaminant wave speed $\hat v = 1+ \delta_7^{1/n}$ when $\delta_7 < 1$. 
The general good agreement provides confidence in the accuracy of the analytical solution.

%To comment: what is really important for application is breakthrough curves. This also motivates use of numerical solution to ensure accuracy of analytical solution. 

%In Fig. \ref{fig:TWsolutions:all} we show the breakthrough curves as obtained by the analytical solutions of the reduced model and the ones obtained from numerically solving the full model.

\subsubsection{Large mass removal and velocity variation}\label{large_mass_discussion}

% \tim{WHEN FRAN ADDS GRAPHS SHOWING HOW RESULTS DIVERGE AS PHI INCREASES THEN WE COULD DISCUSS ...\\
% In the expressions for $Y_{11}$  the first term in the brackets corresponds to the low mass removal result, the second term represents the correction due to the increasing mass of contaminant.
% Defining $f_1(c)$ as the first term and $f_2(c)$ as the second then  in Figure \ref{Phi1Fig} we plot $|f_1]$, $\phi |f_2|$ with $\phi$ varying between 0.1, 0.2, 0.3, 0.4, 0.5. Increasing $\phi$ indicates that the second term becomes increasingly important, however we note that at $c=1/2$ the curves are closest (where both pass through zero. If we look in the region close to this, such that $c = 1/2 -\epsilon$ we find $\lim_{\epsilon \ra 0} |f_2/f_1| = \phi/(2-\phi)$. For this limit to be small then requires $\phi \ll 1$. This isn't as exciting as I had hoped. Implies 25\% error when $\phi=0.4$. Also, I can't do Y12!\\
% Maybe easiest if Fran just adds some comparisons of numerics with increasing phi} 
% \begin{figure*}
%     \centering
% \includegraphics[width=0.5\textwidth]{breakthrough_changing_phi.pdf} 
%          \includegraphics[width=0.5\textwidth]{Y11VaryPhi.png}  
%     \caption{\textcolor{red}{FF: I think we need to recover the picture where evolution of $u$ versus $phi$ is shown. This is what relates large mass removal with velocity variation.} }
%     \label{Phi1Fig}
% \end{figure*}

\begin{figure}
\centering
\begin{subfigure}[c]{0.48\textwidth}
\centering
\includegraphics[width=\textwidth]{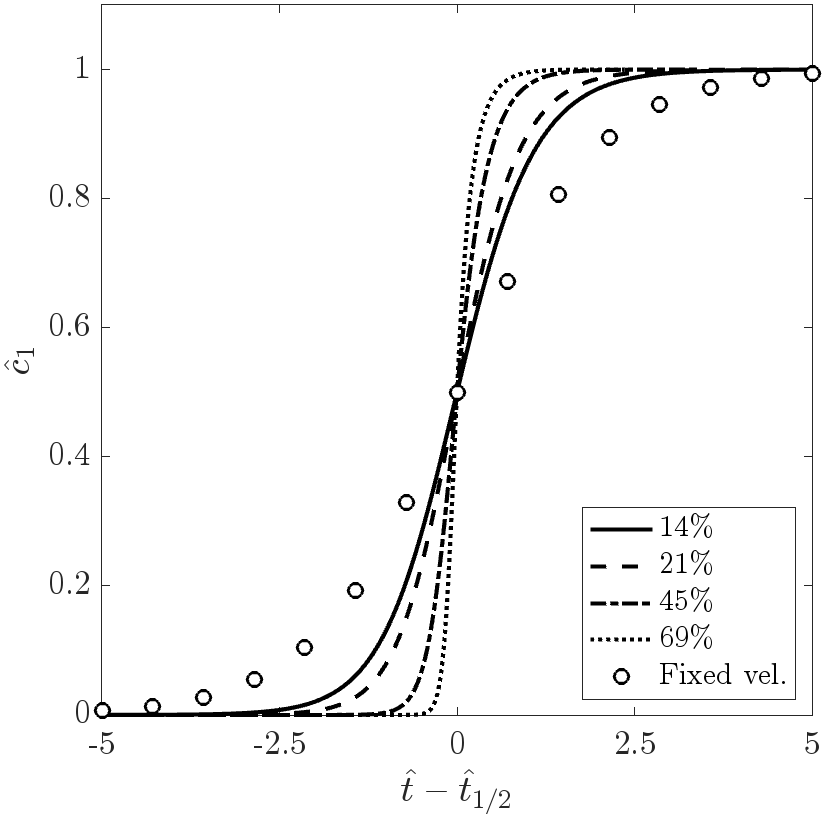}
\caption{}
\label{fig:VariationPhi:a}
\end{subfigure} \quad
\begin{subfigure}[c]{0.48\textwidth}
\centering
\includegraphics[width=\textwidth]{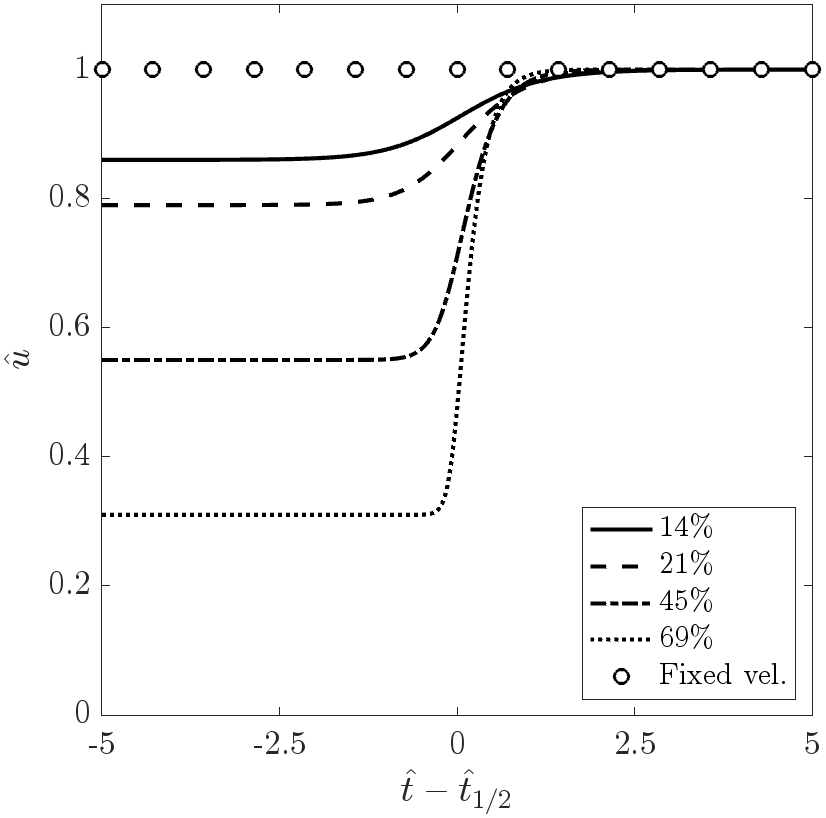}
\caption{}
\label{fig:VariationPhi:b}
\end{subfigure}
\caption{Variation of the contaminant breakthrough curve,  \eqref{BreakthroughModel}, and fluid velocity at the outlet, \eqref{TW:remaining}, for different volume fractions  and $m=n=1$. The fixed velocity curves correspond to setting $\phi_1=0$. The adsorption-to-desorption ratio is $\delta_7=5$.
All parameter values are consistent with the  experiments of  \cite{Delgado2006}, see  Table \ref{tab:Delgadoresults}.} 
\label{fig:VariationPhi}
\end{figure}

The non-dimensional solutions (\ref{TW:remaining}, \ref{BreakthroughModel}) indicate a dependence on only two parameters, $\phi_1, \delta_7$.
In Fig. \ref{fig:VariationPhi} we examine the effect of  varying  $\phi_1$  on the outlet concentration and  velocity, with $\delta_7=5$ fixed. The curves correspond to  the values $\phi_1 = 0.14, 0.21, 0.45, 0.69$, which are taken from the   experiments of \cite{Delgado2006} and
described in \S\ref{sec:AnalyticsVSExperiments}. Also shown is the fixed velocity case, which corresponds to $\phi_1=0$.  We observe that the $\phi_1=0$ case shows a wider spread at breakthrough than the other curves: the slopes decrease in width with increasing $\phi_1$. Note, even at 14\% there is a significant difference from the fixed velocity result (we will discuss this later). In Fig. \ref{fig:VariationPhi:b} we show the evolution of velocity at the outlet. For times sufficiently earlier than  $\hat t_{1/2}$ all contaminant is captured  and the outlet velocity $\hat u(\hat L, \hat t) = 1-\phi_1$ (this may be seen by taking the limit $\hat c_1\to0$ in Eq. \eqref{TW:remaining}).

\section{Application to CO$_\text{2}$ capture}\label{sec:AnalyticsVSExperiments}

In this section, we validate the variable-velocity model against experimental data from Monazam \emph{et al.} \cite{Monazam2013} and Delgado \emph{et al.} \cite{Delgado2006}.

\subsection{Parameter values}\label{sec:parval}

Delgado \emph{et al.} \cite{Delgado2006} provided column breakthrough data for CO$_2$ adsorption onto silicate pellets at four different volumetric percentages of CO$_2$ (14\%, 21\%, 45\% and 69\%, corresponding to their runs 10, 12, 13 and 14). The values of the relevant thermophysical parameters extracted from their original work are listed in Table~\ref{tab:Delgadoproperties1}. The dispersion coefficient has been approximated using the experimental chart of Levenspiel \cite{Levenspiel1999} which relates the inverse Bodenstein number with the Reynolds number of the flow in a packed bed. An estimate of the initial concentrations is obtained from the corresponding temperature and pressure measurements using the ideal gas law. The parameter values for $q_{max}$ and $K$ were found by fitting \eqref{Sipsiso} to the isotherm  data of \cite{Delgado2006}: this calculation is described in the Supplementary Material. Note that the isotherm fitting process also indicates the  kinetics of the adsorption process (physisorption or chemisorption) and, in the case of chemisorption, the order of the reaction. Since the best fit of the data of \cite{Delgado2006} is obtained with $(m,n)=(1,1)$ this mechanism could be either physical or chemical (\textit{i.e.} a chemical reaction with kinetic orders 1). In the Supplementary Material we note that at room  temperature CO$_2$ molecules are adsorbed by Van der Waals forces inside the micropores of the silicalite structure \cite{Grajciar2012,Jiang2020,Santoro2011}, consequently we may assume that the (1,1) behaviour in this case corresponds to physisorption.

\begin{table}[H]
    \centering
    \caption{Material properties and operating conditions from the column adsorption experiments in \cite{Delgado2006}. Each number of run refers to the original number in \cite{Delgado2006}.}
    \label{tab:Delgadoproperties1}
    \begin{tabular}{|l|c|c|c|c|c |c|}\hline
        \cellcolor[rgb]{0.7922,0.8471, 0.8706}\textbf{Property} & \cellcolor[rgb]{0.7922,0.8471, 0.8706}\textbf{Symbol} & \cellcolor[rgb]{0.7922,0.8471, 0.8706}\textbf{Units} & \multicolumn{4}{c|}{\cellcolor[rgb]{0.7922,0.8471, 0.8706}\textbf{Value}} \\ \hline \hline 
        \multicolumn{3}{|c|}{\cellcolor[rgb]{0.7922,0.8471, 0.8706}\textbf{Run}} & \textbf{10} & \textbf{12} & \textbf{13} & \textbf{14} \\ \hline \hline
        CO$_2$ vol. fraction & $\phi_1$ & - & 0.14 & 0.21 & 0.45 & 0.69 \\
        Inlet CO$_2$ conc. & $c_{10}$ & mol/m$^3$ & 5.436 & 8.086 & 17.22 & 26.15 \\
        Inlet He conc. & $c_{20}$ & mol/m$^3$ & 33.392 & 30.419 & 21.045 & 11.749 \\
        Final ad. fraction & $q_{e,exp}$ & mol/kg & 0.28 & 0.38 & 0.73 & 1.08 \\
        Max. ad. fraction & $q_{max}$ & mol/kg & 4.975 & 4.664 & 4.595 & 4.845 \\
        Inlet velocity ($\times$10$^{-3}$) & $u_0$ & m/s & 3.223 & 3.328 & 3.443 & 2.764 \\
        Pressure ($\times$10$^{4}$) & $p_0$ & Pa & 9.62 & 9.54 & 9.48 & 9.39 \\
        Column length & $L$ & m & \multicolumn{4}{c|}{$0.163$} \\
        Column radius & $R$ & m & \multicolumn{4}{c|}{$0.008$} \\
        Void fraction & $\epsilon$ & - & \multicolumn{4}{c|}{$0.52$} \\
        Particle density & $\rho_a$ & kg/m$^3$ & \multicolumn{4}{c|}{$1070$} \\
        Bulk density & $\rho_b$ & kg/m$^3$ & \multicolumn{4}{c|}{$513.6$} \\
        CO$_2$ molar mass & $M_1$ & kg/mol & \multicolumn{4}{c|}{$0.044$} \\
        He molar mass & $M_2$ & kg/mol & \multicolumn{4}{c|}{$0.004$}\\
        Particle diameter & $d_p$ & m & \multicolumn{4}{c|}{$1.4\times10^{-3}$}\\
        Dispersion coef. & $D$ & m$^2$/s & \multicolumn{4}{c|}{$9.03\times10^{-6}$}\\
        Temperature & $T$ & $^\circ$C & \multicolumn{4}{c|}{$25$} \\
        Gas viscosity & $\mu_g$ & mPa$\cdot$s & \multicolumn{4}{c|}{$0.0192$} \\
        Eq. constant & $K$ & m$^3$/mol & \multicolumn{4}{c|}{$0.01096$} \\
        Partial orders & $m,n$ & - & \multicolumn{4}{c|}{1,1} \\         
        \hline
    \end{tabular}
\end{table}

Monazam \emph{et al.} performed column adsorption experiments \cite{Monazam2013} for CO$_2$ capture on PEI modified silica in an N$_2$ stream. The values of the relevant thermophysical parameters extracted from their original work are listed in Table~\ref{tab:Monproperties}. Two different temperatures and two different volumetric percentages of CO$_2$ (16.6\% and 33.3\%) were studied. The dispersion coefficient was obtained using the expression $D=1.44\times 10^{-5}/\epsilon$, see \cite{Myers20}. Since this coefficient is approximate, it has been assumed to be constant with temperature. The inlet concentrations were calculated using the ideal gas law. The value of $q_{max}$, $K$ and the orders $m, n$ were determined from the isotherm data of \cite{Monazam2013iso} for batch adsorption  with the same adsorbate-adsorbent system. The calculations are described in detail in the Supplementary Material. From their data it appears that the adsorption mechanism changes with temperature, 
at $T=40$$^\circ$C the best fit to the isotherm occurs with $n=2$, $m=1$ while $m=n=1$ fits the  $70$$^\circ$C data. This switch in behaviour can be explained  via specific physical/chemical mechanisms (see Supplementary Material). 

\begin{table}[H]
    \centering
    \caption{Operating conditions and  parameter values from \cite{Monazam2013,Monazam2013iso}. Values separated by a dash refer to different volume fractions.}
    \label{tab:Monproperties}
    \begin{tabular}{|l|c|c|c| c|}\hline
        \cellcolor[rgb]{0.7922,0.8471, 0.8706}\textbf{Property} & \cellcolor[rgb]{0.7922,0.8471, 0.8706}\textbf{Symbol} & \cellcolor[rgb]{0.7922,0.8471, 0.8706}\textbf{Units} & \multicolumn{2}{c|}{\cellcolor[rgb]{0.7922,0.8471, 0.8706}\textbf{Value}} \\ \hline \hline
        Temperature & $T$ & $^\circ$C & $40$ & $70$ \\
        CO$_2$ volume fraction & $\phi_1$ & - & $0.333$ & $0.166$ - $0.333$ \\
        Inlet CO$_2$ conc. & $c_{10}$ & mol/m$^3$ & $12.96$ & $5.90$ - $11.83$ \\
        Inlet N$_2$ conc. & $c_{20}$ & mol/m$^3$ & $25.96$ & $29.62$ - $23.69$ \\
        Final ad. frac. & $q_{e,\text{exp}}$ & mol/kg & $1.99$ & $2.743$ - $2.761$ \\
        Partial orders & $m,n$ & - & 1, 2 & 1, 1 \\
        Max. ad. frac. & $q_{max}$ & mol/kg & $2.37$ & $3.04$ - $2.91$ \\
        Eq. constant & $K$ & m$^{3m}$kg$^{n-1}$mol$^{1-m-n}$ & 2.13 & 1.59 \\
        Gas viscosity & $\mu_g$ & mPa$\cdot$s & $0.0185$ & $0.0198$ \\
        Column length & $L$ & m & \multicolumn{2}{c|}{$0.254$}\\
        Column radius & $R$ & m & \multicolumn{2}{c|}{$0.07$}\\
        Void fraction & $\epsilon$ & - & \multicolumn{2}{c|}{$0.3$}\\
        Particle density & $\rho_a$ & kg/m$^3$ & \multicolumn{2}{c|}{$900$}\\
        Bulk density & $\rho_b$ & kg/m$^3$ & \multicolumn{2}{c|}{$630$}\\
        CO$_2$ molar mass & $M_1$ & kg/mol & \multicolumn{2}{c|}{$0.044$}\\
        N$_2$ molar mass & $M_2$ & kg/mol & \multicolumn{2}{c|}{$0.028$}\\
        Particle diameter & $d_p$ & m & \multicolumn{2}{c|}{$1.5\times10^{-4}$}\\
        Dispersion coef. & $D$ & m$^2$/s & \multicolumn{2}{c|}{$4.8\times10^{-5}$}\\
        In. fluid velocity & $u_0$ & m/s & \multicolumn{2}{c|}{$0.0944$}\\
        Pressure & $p_a$ & Pa (Atm) & \multicolumn{2}{c|}{$101325$ ($1$)}\\
        \hline
    \end{tabular}
\end{table}

\subsection{Results of the fitting to the breakthrough data}

%The models presented in previous sections have been tested against the breakthrough data reported by Monazam \emph{et al.} \cite{Monazam2013} and Delgado \emph{et al.} \cite{Delgado2006}. The fitted experimental data accounts for the concentration profile obtained at the exit of the adsorption column. Regardless of the orders $m$ and $n$, the only fitting parameter is the adsorption coefficient $k_a$, since the rest of parameters have been previously obtained from the original papers \cite{Monazam2013,Delgado2006} or the isotherm analysis in section \ref{sec:parval}. 

The analysis of the isotherm data in the Supplementary Material determines the parameters $q_{max}$, $K$ and the orders $m$, $n$, so the only remaining unknown is the adsorption coefficient $k_a$. Evaluating \eqref{Soln:c1} at $x=L$ yields the breakthrough curves
\begin{equation}
\label{Y11Fit}
  Y_{11} (c_{1b}(t)) = \frac{1+\delta_7}{\mathcal{T}}( t-t_{1/2}) 
\end{equation}
\begin{equation}
\label{Y12Fit}
  Y_{12} (c_{1b}(t)) = \frac{1+\sqrt{\delta_7}}{\mathcal{T}}( t-t_{1/2}) 
\end{equation}
for $(m,n)=(1,1)$ (Langmuir model) and $(m,n)=(1,2)$, respectively. The functions $Y_{11}$, $Y_{12}$ are defined by \eqref{Y11Dim}-\eqref{Y12Dim} and the value of $\delta_7= 1/(K c_{10})$ is known. Note the breakthrough curves \eqref{Y11Fit}-\eqref{Y12Fit} are linear functions in $t-t_{1/2}$ where the slope depends on the unknown $k_a$ through $\mathcal{T}=1/(k_a c_{10} q_{max}^{n-1})$. This permits a straightforward fitting procedure to determine $k_a$ based on measuring how the experimental values of $Y_{mn}(c_{1b}(t))$ deviate from the theoretical linear profile. The fitting procedure is an extension of the one used in \cite{Aguareles23} for the constant velocity model.

%After replacing $\mathcal{T}=1/\left(k_a c_{10}\right)$, $\delta_7 = k_d/(k_a c_{10})$ in 
%equation \eqref{BC11} we may write the implicit expression for the Langmuir model, $(m,n)=(1,1)$, breakthrough curve 
%\begin{equation}
%\label{Y11Fit}
%  Y_{11} (c_{1b}(t)) = \frac{1+\delta_7}{\mathcal{T}}( t-t_{1/2}) = (k_d + k_a c_{10})( t-t_{1/2})
%\end{equation}
%where $Y_{11}$ is defined by \eqref{Y11Dim}. 
%For the Sips model with $(m,n)=(1,2)$, $\mathcal{T}=1/\left(k_a c_{10}q_{max}\right)$, $\delta_7 = k_d/(k_a c_{10})$ and we may express the breakthrough curve as 
%\begin{equation}
%\label{Y12Fit}
%  Y_{12} (c_{1b}(t)) = \frac{1+\sqrt{\delta_7}}{\mathcal{T}}( t-t_{1/2}) = k_a c_{10}q_{max} \left(1+\sqrt{\frac{k_d}{k_a c_{10}}}\right)( t-t_{1/2})
%\end{equation}
%where $Y_{12}$ is defined by \eqref{Y12Dim}. Although the above forms are not as intuitive as writing out $Y_{mn}$ explicitly and, at least in the first case, rearranging to the form $c_{1b}(t)$ they have a distinct advantage. Calculating $Y_{mn}$ by inputting the experimental values for $c_{1b}$ and then plotting this against the experimental values for $t-t_{1/2}$ will produce a straight line passing through the origin with the correct model, where the slope is determined by adjusting $\mathcal{T}$ or $k_a$. This permits a robust fitting as opposed to in the $c, t$ domain.

%Since the value $K=k_a/k_d$ is determined from the isotherm the only unknown in the above equations is $\mathcal{T}$ or $k_a$. 

\begin{figure}[H]
\centering
\begin{subfigure}[c]{0.48\textwidth}
\centering
\includegraphics[width=\textwidth]{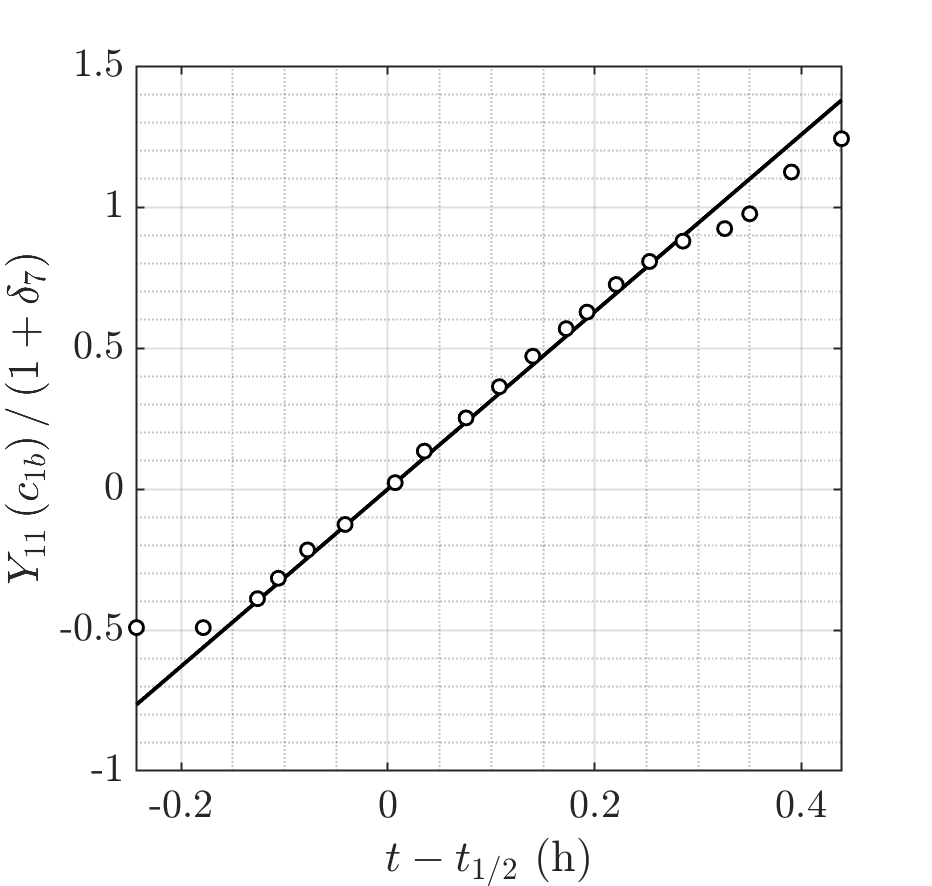}
\caption{}
\label{DelFit:a}
\end{subfigure} \quad
\begin{subfigure}[c]{0.48\textwidth}
\centering
\includegraphics[width=\textwidth]{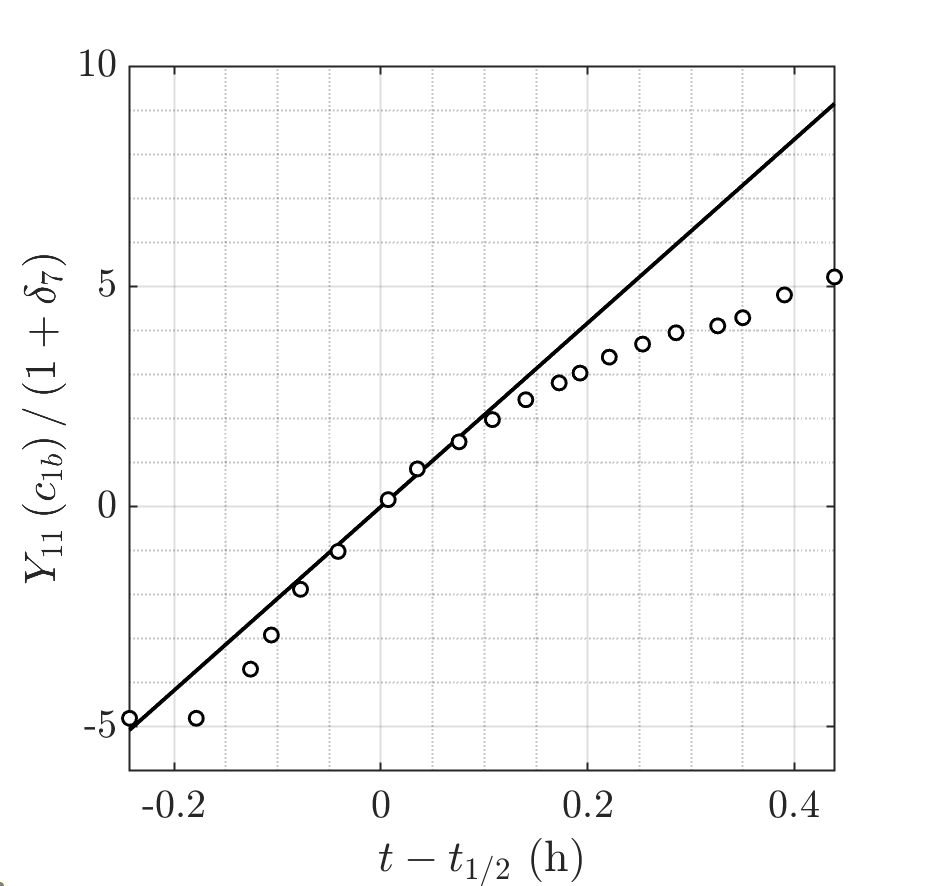}
\caption{}
\label{DelFit:b}
\end{subfigure}
\caption{Dataset 14 of Delgado \textit{et al.} \cite{Delgado2006}, where the experimental data points are shown in the form $Y_{11}\left(c_{1b}\right)/(1+\delta_7)$ against time $t-t_{1/2}$. The solid line is a straight line passing through the origin with a slope $1/\mathcal{T}$. On panel (a), $Y_{11}$ is calculated for the variable velocity case with $\phi=0.69$. On panel (b), $Y_{11}$ is calculated for the constant velocity case, obtained by setting $\phi = 0$.} 
\label{DelFit}
\end{figure}

Noting that the adsorption process studied in Delgado \textit{et al.} \cite{Delgado2006} corresponds to a $(1,1)$ model in Fig. \ref{DelFit:a} we compare the predictions of equation \eqref{Y11Fit} with the experimental Dataset 14,  which has a volume fraction of CO$_2$ of 69\% ($\phi_1=0.69$). The data points on the graph are generated by plotting $Y_{11}(c_{1b})/(1+\delta_7)$ using the experimental  values of $c_{1b}(t)$ versus the times $t-t_{1/2}$, where $t_{1/2} = 2314$s = 0.642 hours. The solid line corresponds to the linear fit obtained using the \texttt{fit} function from the Matlab curve fitting toolbox. Fig. \ref{DelFit:b} shows the prediction using the constant velocity model (i.e., equation \eqref{Y11Fit} with $\phi=0$).  Comparison of  Figs. \ref{DelFit:a} and \ref{DelFit:b} demonstrate that the variable velocity model provides a much more accurate fit to the experimental data than the constant velocity model, which clearly deviates from the linear profile. 

Repeating the fitting for all Runs  detailed in Table \ref{tab:Delgadoproperties1} we obtain the results shown in Table \ref{tab:Delgadoresultsgof}.
For Run 14 we observe that the variable velocity model has an $R^2=0.994$ and SSE of 0.01 while the values for the constant velocity model are 0.94, 4.8. Since the SSE refers to a logarithmic function of the experimental data its size is not so important, rather the important feature is that the variable velocity value is nearly five hundred times smaller than the constant velocity value. Although it is obvious from the graphs, Table \ref{tab:Delgadoresultsgof} provides quantitative evidence that the constant velocity model is  the most accurate when $\phi_1=0.69$. Reducing the volume fraction reduces the difference between the results: for the lowest value $\phi_1 = 0.14$ the SSE is only a factor 17 less than the $\phi_1=0$ result. In all cases the variable velocity model has a lower SSE and higher $R^2$ than the constant velocity result.

%Delgado results, 1st comparison and then parameters%%%%%%%%%%%%%%%%%%%%

%The case of study of Delgado \emph{et al.} \cite{Delgado2006} clearly shows that the kinetics of adsorption account for a ratio $m/n=1$, as seen in Fig.~\ref{fig:Delgadoisotherm} and Table~\ref{tab:Tableiso}.  Thus, the $(1,2)$ model has not been considered for fitting the breakthrough data, and only the $(1,1)$ model \eqref{BC11} (here variable velocity) and the constant velocity model (equation \eqref{BC11}) are considered. The fitted parameter $k_a$ as well as parameters that assess the goodness of the fit are reported in Table~\ref{tab:Delgadoresultsgof}.

\begin{table}[H]
    \centering
    \caption{Comparison between the fitting to Delgado \emph{et al.} \cite{Delgado2006} breakthrough data of the $(m,n)=(1,1)$ variable velocity model (equation \eqref{BC11}) and the $(m,n)=(1,1)$ constant velocity model.}
    \label{tab:Delgadoresultsgof}
    \begin{tabular}{|c|c|c|c|c|c|}
    \hline
      \cellcolor[rgb]{0.7922,0.8471, 0.8706}\textbf{Parameter} & \multicolumn{4}{c|}{\cellcolor[rgb]{0.7922,0.8471, 0.8706}\textbf{Value}}  \\ \hline \hline
      \cellcolor[rgb]{0.7922,0.8471, 0.8706}\textbf{Run} & \textbf{10} & \textbf{12} & \textbf{13} & \textbf{14}  \\ \hline
      $\phi_1$ (-) & 0.14 & 0.21 & 0.45 & 0.69 \\ \hline\hline
      \multicolumn{5}{|c|}{\cellcolor[rgb]{0.7922,0.8471, 0.8706}\textbf{Variable velocity}} \\ \hline
      $k_a$ ($\times 10^{-4}$) (m$^3$mol$^{-1}$s$^{-1}$) & 1.341 & 0.971 & 0.592 & 0.334 \\ \hline
      SSE & 0.667 & 1.251 & 0.102 & 0.010 \\ \hline
      R-squared & 0.9433 & 0.9110 & 0.9746 & 0.9942 \\ \hline\hline    
      \multicolumn{5}{|c|}{\cellcolor[rgb]{0.7922,0.8471, 0.8706}\textbf{Constant velocity}} \\ \hline
      $k_a$ ($\times 10^{-3}$) (m$^3$mol$^{-1}$s$^{-1}$) & 0.472 & 0.374 & 0.268 & 0.221 \\ \hline
      SSE & 11.375 & 28.540 & 9.259 & 4.800 \\ \hline
      R-squared & 0.9248 & 0.8681 & 0.8993 & 0.9395 \\ \hline
    \end{tabular}
\end{table}

%Note that the fittings shown in Figure \ref{DelFit} correspond to Run 14 in Table \ref{tab:Delgadoresultsgof}. All cases in this table show a better agreement with the variable velocity model \eqref{BC11} than with the constant velocity one. In some cases the better agreement of the variable velocity model can be very significant compared to the constant velocity. This is the case of run 14 (69\% v/v) for instance, since the SSE is two orders of magnitude higher than with variable velocity (4.800 compared to 0.010, respectively). Taking into account these results, the rest of parameters are provided in Table~\ref{tab:Delgadoresults}.

\begin{table}[H]
    \centering
    \caption{Results obtained from the fitting to Delgado \emph{et al.} \cite{Delgado2006} breakthrough data of the model with $(m,n)=(1,1)$ (equation \eqref{BC11}). The only fitting parameter is $k_a$, the rest are obtained using the values in Table~\ref{tab:Delgadoproperties1}.}
    \label{tab:Delgadoresults}
    \begin{tabular}{|c|c|c|c|c|}
    \hline
      \cellcolor[rgb]{0.7922,0.8471, 0.8706}\textbf{Parameter} & \multicolumn{4}{c|}{\cellcolor[rgb]{0.7922,0.8471, 0.8706}\textbf{Value}}  \\ \hline \hline
      \cellcolor[rgb]{0.7922,0.8471, 0.8706}\textbf{Run} & \textbf{10} & \textbf{12} & \textbf{13} & \textbf{14}  \\ \hline \hline
      $\phi_1$ (-) & 0.14 & 0.21 & 0.45 & 0.69 \\ \hline  
      $k_a$ ($\times 10^{-4}$) (m$^3$mol$^{-1}$s$^{-1}$) & 1.341 & 0.971 & 0.592 & 0.334 \\ \hline
      $k_d$ ($\times 10^{-2}$) (s$^{-1}$) & 1.223 & 0.886 & 0.539 & 0.304 \\ \hline
      $\mathcal{T}$ (s) & 1371.62 & 1272.97 & 981.74 & 1145.54 \\ \hline
      $\mathcal{L}$ ($\times 10^{-2}$) (m) & 0.4891 & 0.7438 & 1.283 & 1.731 \\ \hline
      $\mathcal{P}$ ($\times 10^{-2}$) (Pa) & 1.893 & 2.937 & 4.936 & 5.055 \\ \hline
      $\delta_1$ ($\times 10^{-3}$) & 1.106 & 1.755 & 3.794 & 5.465 \\ \hline
      $\delta_2$ & 0.573 & 0.376 & 0.218 & 0.162 \\ \hline
      $\delta_3$ ($\times 10^{-7}$) & 1.968 & 3.079 & 5.207 & 5.383 \\ \hline
      $\delta_4$ ($\times 10^{-3}$) & 4.137 & 3.938 & 2.992 & 1.418 \\ \hline
      $\delta_5$ & 1.791 & 2.924 & 9.000 & 24.48 \\ \hline
      $\delta_6$ ($\times 10^{-4}$) & 1.313 & 1.342 & 1.820 & 1.739 \\ \hline
      $\delta_7$ & 16.769 & 11.273 & 5.294 & 3.486 \\ \hline
    \end{tabular}
\end{table}

The analytical solutions presented in Section~\ref{sec:AnalyticalSolns} were derived assuming that some of the dimensionless parameters of the model are typically small in column adsorption processes. In Table~\ref{tab:Delgadoresults} we show the value of these parameters along with $k_a$, $k_d$ and the resulting values of the time, length, and pressure scales, for the data of \cite{Delgado2006}. The orders of magnitude of $\delta_1$, $\delta_3$, $\delta_4$ and $\delta_6$ are very small (between 10$^{-7}$ and 10$^{-3}$). Although $\delta_5$ is $\ord{10}$, the product $\delta_4\delta_5=\ord{10^{-2}}$, which is consistent with the model assumptions. In this example the column is working at a relatively low concentration for the adsorbent capacity. Thus, $\delta_7$ is high, meaning that desorption is significant for most of the process. The value of $\delta_2$ is relatively high because of the low velocity of the process. However, it is still low enough for the assumptions of the model to hold. 

In previous analytical  work on the removal of trace amounts it has been stressed that 
the adsorption coefficient, $k_a$,  must remain constant with respect to concentration, \cite{Crol23, Myers23, Myers24}, otherwise the solution method is invalid. However, it is well-known that $k_a$ may vary with quantities such as temperature, pressure or flow rate. Here, the removal of large quantities of the fluid results in variations in both pressure and velocity. The effect of the velocity  $u$ on mass transfer parameters has previously been  reported in the literature, see \cite{Puiggene1997}, prompting various experimental correlations that define $k_a$ as a function  of $u$. In Table~\ref{tab:Delgadoresults} the reduction in $k_a$ is approximately proportional to the reduction of $1-\phi_1$, which coincides with the change in velocity (between $[1-\phi_1, 1]$) 
suggesting the possibility of a linear variation $k_a \propto u$. We also note that as $\phi_1 \ra 0$ the velocity variation tends to zero, which is then consistent with a fixed $k_a$ for the removal of trace amounts.

\begin{figure}[H]
    \centering
    \begin{subfigure}[b]{0.48\textwidth}
         \centering
         \includegraphics[width=\textwidth]{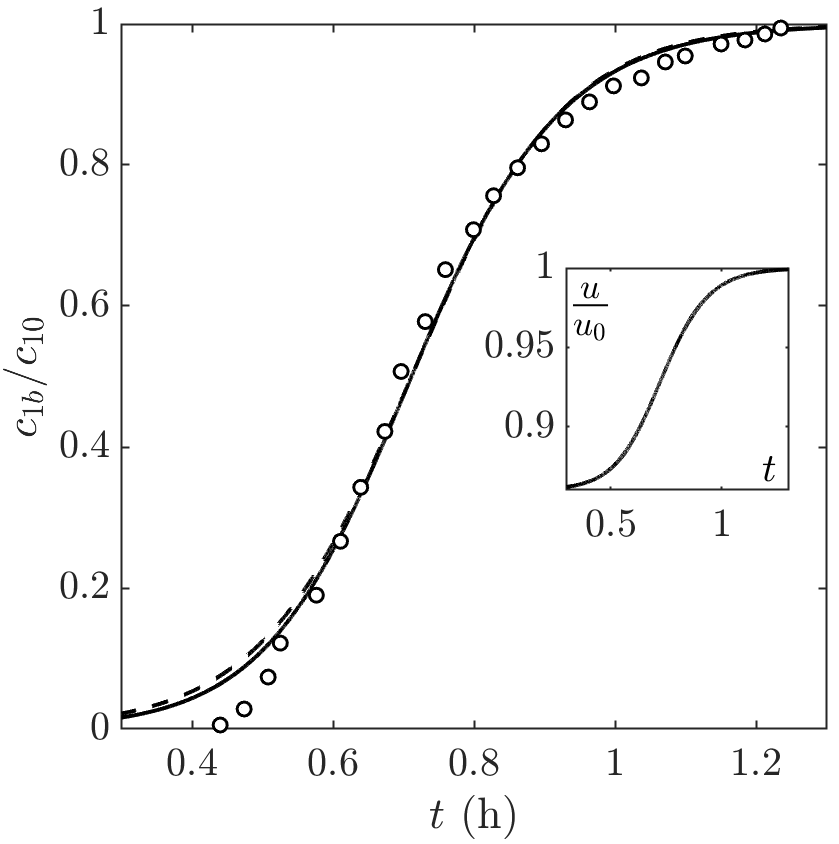}
         \caption{14\% CO$_2$.}
         \label{fig:TWsolutions:Delgadoc0:14}
     \end{subfigure}
     \hfill
     \begin{subfigure}[b]{0.48\textwidth}
         \centering
         \includegraphics[width=\textwidth]{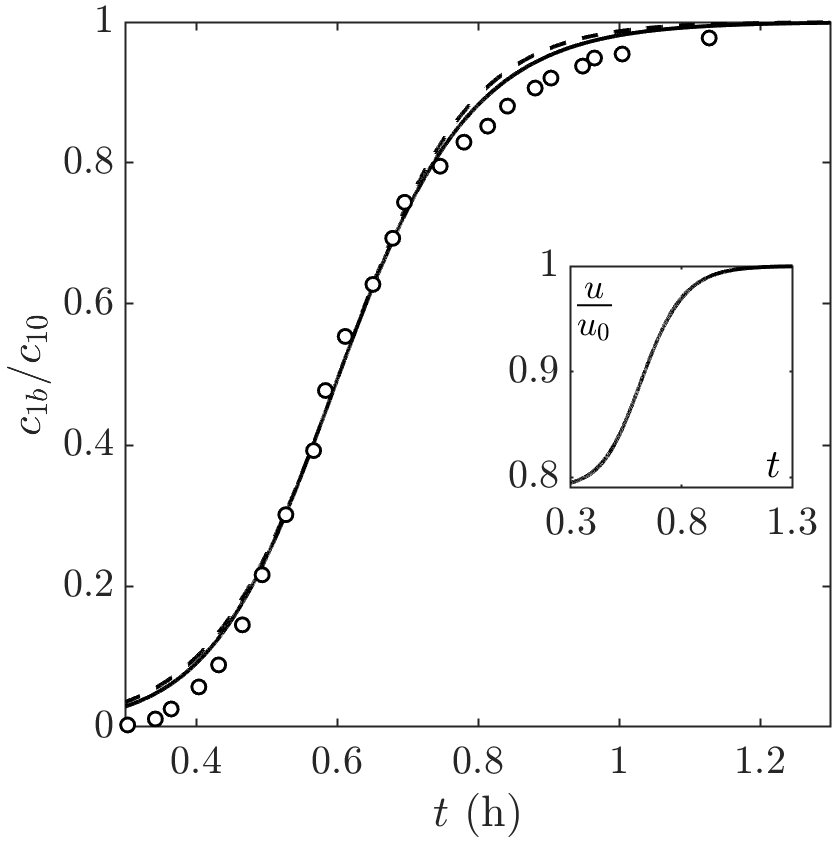}
         \caption{21\% CO$_2$.}
         \label{fig:TWsolutions:Delgadoc0:21}
     \end{subfigure}
     \\
     \begin{subfigure}[b]{0.48\textwidth}
         \centering
         \includegraphics[width=\textwidth]{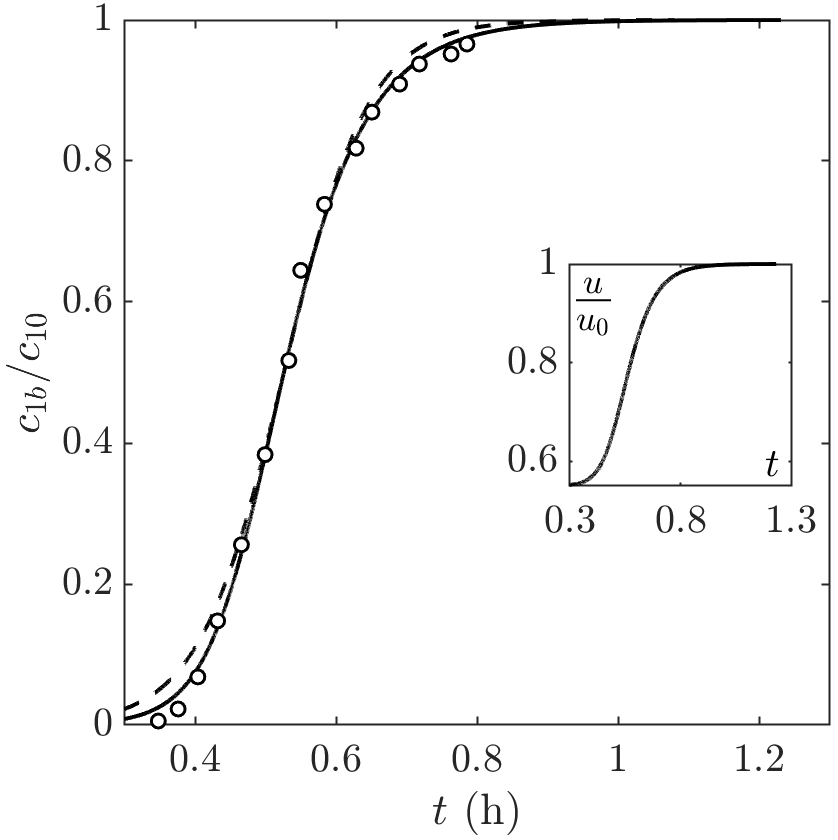}
         \caption{45\% CO$_2$.}
         \label{fig:TWsolutions:Delgadoc0:45}
     \end{subfigure}
     \hfill
     \begin{subfigure}[b]{0.48\textwidth}
         \centering
         \includegraphics[width=\textwidth]{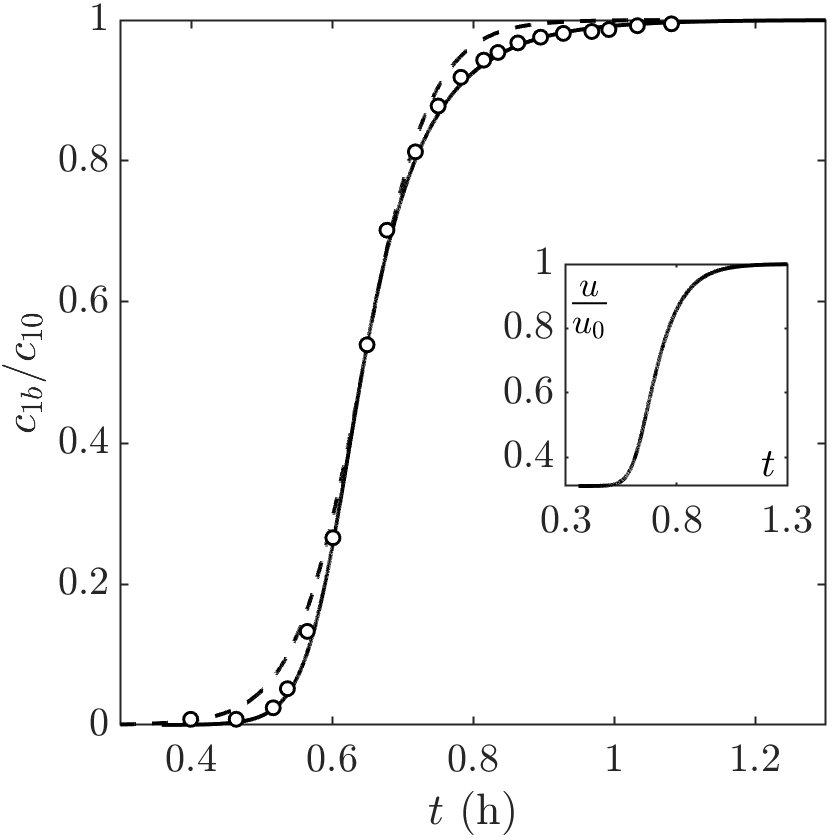}
         \caption{69\% CO$_2$.}
         \label{fig:TWsolutions:Delgadoc0:69}
     \end{subfigure}
    \caption{Fitting of the model accounting for $(m,n)=(1,1)$ to the breakthrough data reported by Delgado \emph{et al.} \cite{Delgado2006} with a range of CO$_2$ volumetric percentages $\phi_1$ at the inlet of the column (runs 10, 12, 13 and 14 in Table~\ref{tab:Delgadoproperties1}). The constant velocity model (dashed line) is retrieved when setting $\phi_{1}\rightarrow 0$ in Eq.~\eqref{BC11}. In each figure, the inset shows the evolution of the flow velocity at the outlet as per Eq.~\eqref{Soln:others}.}
    \label{fig:Delgadoc0}
\end{figure}

Whilst the linear form is an excellent way to understand the experimental data, the more standard procedure is to plot $c_b(t)$. In Figure \ref{fig:Delgadoc0} we plot the breakthrough data for the four cases of Runs 10, 12, 13, 14 and compare with the constant velocity model. A key observation here is that in this format, for the low concentration cases $\phi_1 = 0.14, 0.21$ the differences between the variable and constant velocity results are virtually indistinguishable. This indicates that if we carry out a fitting based on this format it would be difficult to identify the correct model. From the linear format of Figure \ref{DelFit} it is clear that the variable velocity model is the correct one and therefore the value of $k_a$ calculated by this method should be the more accurate. Both models miss a number of points, particularly when $\phi_1 = 0.21$, this could be due to some experimental error. For the high concentration results the variable velocity model captures almost all data points exactly while the constant velocity model reduces in accuracy.

We stress that for the low concentration results the close proximity of the variable and constant velocity curves does not imply that the results are equivalent. Rather it demonstrates that with parameter values adjusted appropriately the incorrect model may appear to be a good approximation. This may also be understood through
Figure \ref{fig:VariationPhi} which shows that, with all other quantities held constant the non-dimensional breakthrough curves vary significantly between the $\phi_1=0, 0.14$ cases. The difference in the form of the breakthrough curves is due to the second term in square brackets of equation (\ref{TWsolution:Yc}a) which is proportional to $\phi_1$. The difference therefore increases with increasing $\phi_1$, however the scale of the difference changes due to the factor $1/(1+\phi_1 \delta_7)$ outside the brackets, which will magnify differences for small $\phi_1$. In dimensional form, equation \eqref{BC11} introduces an additional factor $\tau = 1/(k_a c_{10})$, so any magnification of differences can be reduced by choosing a higher value of $k_a$ (we note that in Tables \ref{tab:Delgadoresultsgof}, \ref{tab:Delgadoresults}  $k_a$ decreases with increasing $\phi_1$). Consequently, the standard form of presenting breakthrough results may be a poor choice for fitting since it permits the incorrect model to appear accurate
and the calculated value of $k_a$  will not reflect the correct value of the adsorbent-adsorbate system. 
The linear form appears to be a more robust and accurate method.

\begin{figure}[H]
\centering
\begin{subfigure}[c]{0.48\textwidth}
\centering
\includegraphics[width=\textwidth]{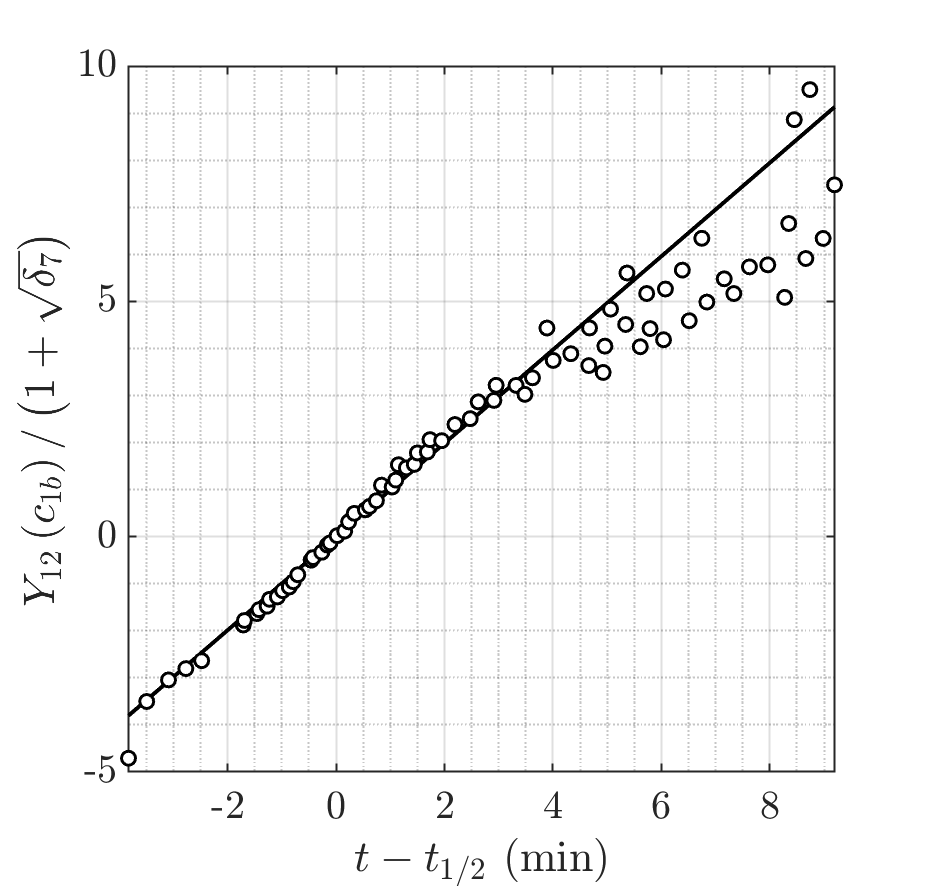}
\caption{}
\label{Monfit:a}
\end{subfigure} 
\begin{subfigure}[c]{0.48\textwidth}
\centering
\includegraphics[width=\textwidth]{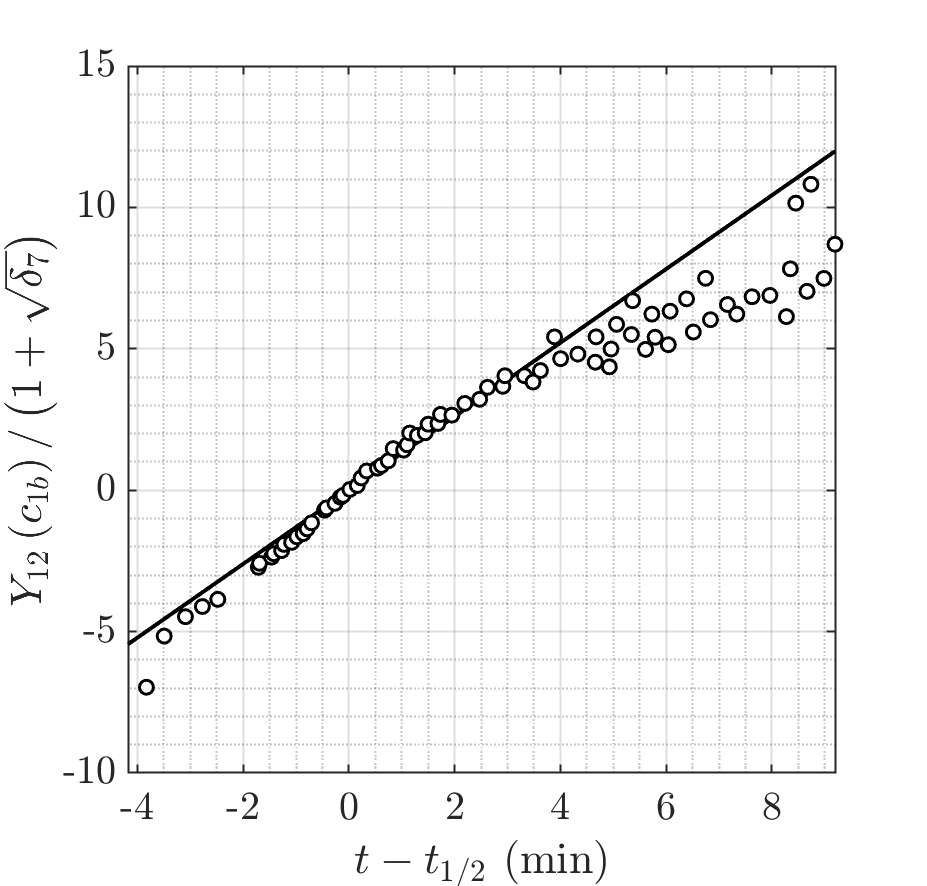}
\caption{}
\label{Monfit:b}
\end{subfigure} 
\caption{Linear fitting of Monazam \textit{et al.} \cite{Monazam2013} data at 40$^\circ$C using equation \eqref{Y12Fit}. On panel (a), $Y_{12}$ is calculated for the variable velocity case with $\phi_1=0.333$. On panel (b), $Y_{12}$ is calculated for the constant velocity case, $\phi_1 = 0$. The value of $\mathcal{T}$ is retrieved from the inverse of the slope of the line.} 
\label{Monfit}
\end{figure}

We now analyse the data of Monazam \textit{et al.} \cite{Monazam2013} for $T=40\degree{C}$ which is shown in the Supplementary Material to follow a $(m,n)=(1,2)$ chemisorption model. In Figure \ref{Monfit:a} we compare the prediction of equation \eqref{Y12Fit} with the experimental data. The data points are generated by plotting $Y_{12}(c_{1b})/(1+\sqrt{\delta_7})$ computed with the outlet experimental values of $c_{1b}(t)$ against $t-t_{1/2}$. The linear fit is obtained by using the \texttt{fit} function from the Matlab curve fitting toolbox. In Figure \ref{Monfit:b} we show the analogous results obtained with the constant velocity model (setting $\phi_1=0$ in \eqref{Y12Fit}). The variable velocity model captures the majority of data points at early times while the constant velocity model shows a poor agreement in this region. For large times the data is more scattered and both models give a relatively poor fit, however the deviations are clearly larger in the constant velocity model.

%In Figure \ref{fig:Mongbt} we show the results of Figure \ref{Monfit} in the standard format of $c_{1b}(t)$ using equation \eqref{BC12} with $\phi_1=0.333$ and $\phi=0$ (constant velocity), which correspond to $\mathcal{T} =1.68 \times 10^{-2}$ s$^{-1}$ and $\mathcal{T} =1.28 \times 10^{-2}$ s$^{-1}$, respectively. As in \ref{Monfit}, it is obvious that the $(1,2)$ model provides a close fit to the experimental data. \textcolor{blue}{Since $\phi_1$ is significantly lower than the one reported by Delgado \textit{et al.} \cite{Delgado2006} for the fitting in Figure \ref{DelFit} ($\phi_1=0.69$), the differences between the variable and constant velocity models are more subtle. Despite these differences are almost unnoticed in the standard form (see Figure \ref{fig:Mongbt}), the linear fitting of the constant velocity model (Figure \ref{Monfit:b}) looses the points at early and later times. The variable velocity model (Figure \ref{Monfit:a}) instead is capable to capture most of the points at early times as well as the points with higher concentrations at late times. Note that the noise in these last data points gives the impression of a poor agreement with the model. However, this disagreement is almost not noticeable in the standard form (Figure \ref{Mongbt:b}). This indicates how sensitive the linear plot is to deviations far from the half time.

The results of the fitting process for Figure \ref{Monfit} are shown in the first column of Table~\ref{tab:Monresultsgof}. While the $R^2$ values both indicate a good fit, the SSE value is three times higher for the constant velocity model. We also show the results of fitting to the 70$^\circ$C data, which is modelled by a (1,1) reaction or physisorption. Again the $R^2$ values are close and for the case of $\phi_1 = 0.166$ the SSEs have a similar magnitude. 

\begin{figure}[H]
    \centering
    \includegraphics[width=.68\textwidth]{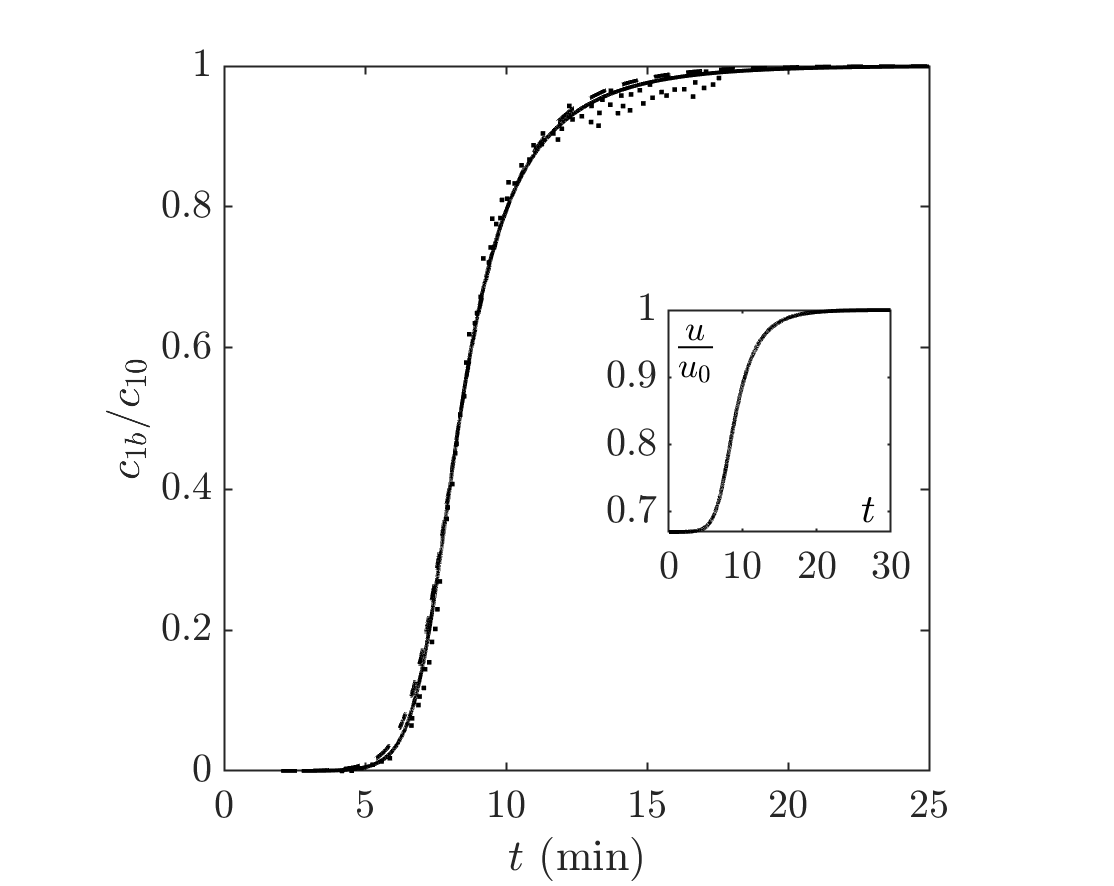}
    \caption{The dots correspond to the breakthrough data from Monazam \textit{et al.} \cite{Monazam2013} at 40$^\circ$C. The solid and dashed lines  are the breakthrough curves from the variable and constant velocity models, respectively, for $(m,n)=(1,2)$ with $k_a$ obtained from the linear fit (see Table~\ref{tab:Monresultsgof}).
    %Standard format of the fitting of breakthrough data by Monazam \textit{et al.} \cite{Monazam2013} shown in Figure \ref{Monfit}, at 40$^\circ$C, $u_0=0.0944$ m/s and $\phi_1=0.333$. The variable velocity model (solid line) is compared with the constant velocity model (dashed line) for $(m,n)=(1,2)$. The inset shows the evolution of the flow velocity at the outlet for $(m,n)=(1,2)$ as per Eq.~\eqref{Soln:others}.
    } 
    \label{fig:Mongbt}
\end{figure}

In Figure \ref{Monfit} it is apparent that the variable velocity model is more accurate for the case $T=$
40$^\circ$C data,  $\phi_1 = 0.333$.
However, when plotting the standard $c_{b}(t)$  form of the breakthrough, shown in  Figure \ref{fig:Mongbt}, there is no noticeable difference between the two curves, although the $k_a$ predicted for the constant velocity case is 30\% higher than the variable velocity one. This confirms our assertion  that fitting to the linear form is more robust than to the standard form of the breakthrough curve.

\begin{table}[H]
    \centering
    \caption{Results of the fitting procedure for Monazam \emph{et al.} \cite{Monazam2013} breakthrough data at 40$^\circ$C and 70$^\circ$C. 
    %of the models with $(m,n)=(1,1)$ (equation \eqref{BC11}) and $(1,2)$ (equation \eqref{BC12}), and the $(1,2)$ constant velocity model for 40$^\circ$C, and $(1,1)$ constant velocity model for 70$^\circ$C. The constant velocity models are retrieved when setting $\phi_{1}\rightarrow 0$ in Eqs.~\eqref{BC11} and \eqref{BC12}.
    }
    \label{tab:Monresultsgof}
    \begin{tabular}{|c|c|c|c|c|c|c|}
    \hline
      \cellcolor[rgb]{0.7922,0.8471, 0.8706}\textbf{Parameter} & \multicolumn{3}{c|}{\cellcolor[rgb]{0.7922,0.8471, 0.8706}\textbf{Value}}  \\ \hline \hline
      $\phi_1$ (-) & \multicolumn{2}{c|}{$0.333$} & $0.166$ \\ \hline
      $T$ ($^\circ$C) & $40$ & \multicolumn{2}{c|}{$70$} \\ \hline
      Partial orders & (1,2) & \multicolumn{2}{c|}{(1,1)}\\ \hline\hline
      \multicolumn{4}{|c|}{\cellcolor[rgb]{0.7922,0.8471, 0.8706}\textbf{Variable velocity ($\phi_1>0$)}} \\ \hline
      $k_a$ ($\times 10^{-4}$) (m$^3$kg$^{n-1}$mol$^{-n}$s$^{-1}$) & 5.388 & 6.230 & 7.088 \\ \hline
      SSE & 2.865 & 1.297 & 27.412 \\ \hline
      R-squared & 0.9857 & 0.9831 & 0.9081 \\ \hline\hline
      \multicolumn{4}{|c|}{\cellcolor[rgb]{0.7922,0.8471, 0.8706}\textbf{Constant velocity ($\phi_1=0$)}} \\ \hline
      $k_a$ ($\times 10^{-4}$) (m$^3$kg$^{n-1}$mol$^{-n}$s$^{-1}$) & 7.074 & 7.392 & 7.673 \\ \hline
      SSE & 8.482 & 2.882 & 31.126 \\ \hline
      R-squared & 0.9762 & 0.9746 & 0.9130 \\ \hline
    \end{tabular}
\end{table}

\begin{figure}[H]
    \centering
    \begin{subfigure}[b]{0.48\textwidth}
         \centering
         \includegraphics[width=\textwidth]{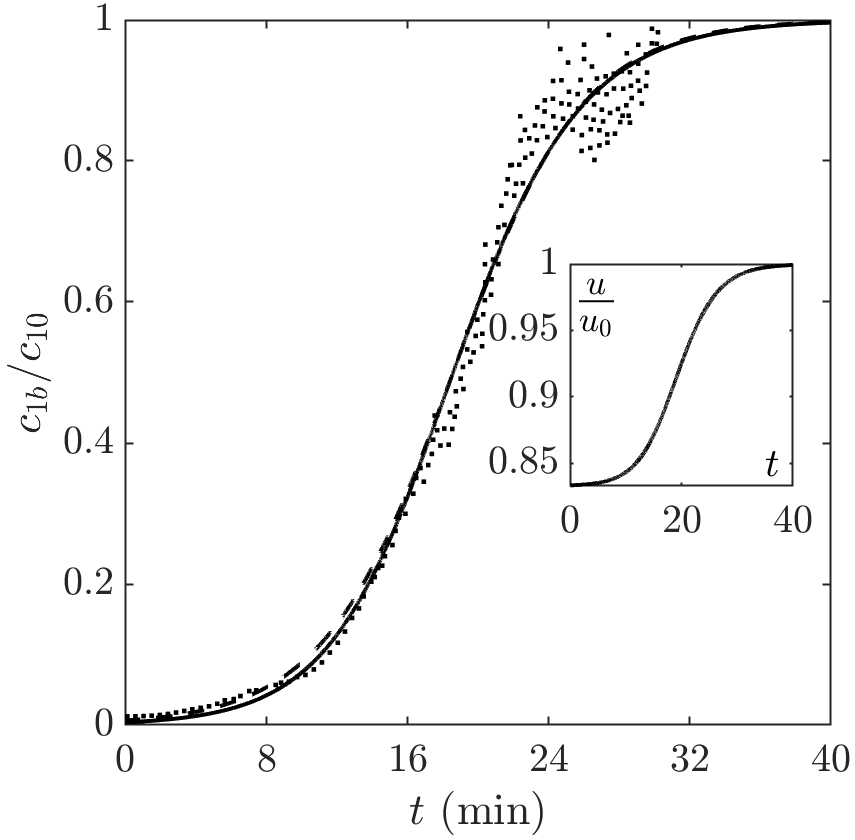}
         \caption{16.6\% CO$_2$.}
         \label{fig:TWsolutions:Monc0:16}
     \end{subfigure}
     \hfill
     \begin{subfigure}[b]{0.48\textwidth}
         \centering
         \includegraphics[width=\textwidth]{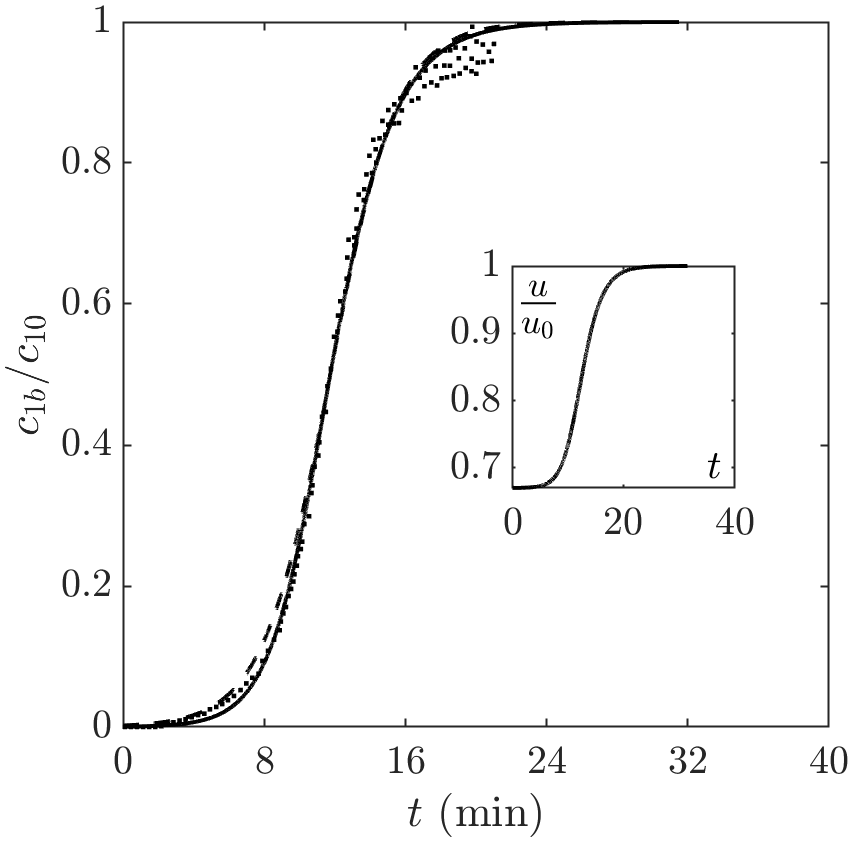}
         \caption{33.3\% CO$_2$.}
         \label{fig:TWsolutions:Monc0:33}
     \end{subfigure}
    \caption{Fitting of the model to the breakthrough data reported by Monazam \emph{et al.} \cite{Monazam2013} at 30 standard L/min and 70 $^\circ$C with different volumetric percentages at the inlet of the column. Solid line: $(1,1)$ variable velocity model (equation \eqref{BC11}); dashed line: $(1,1)$ constant velocity model. 
    The constant velocity model is retrieved when setting $\phi_{1}\rightarrow 0$ in Eq.~\eqref{BC11}.  In each figure, the inset shows the evolution of the flow velocity at the outlet as per Eq.~\eqref{Soln:others}.}
    \label{fig:Monc0}
\end{figure}

In Figure~\ref{fig:Monc0} we present the breakthrough curves for Monazam's data at 70$^\circ$C. As suggested by the SSE and R$^2$ values in Table~\ref{tab:Monresultsgof}, the fitting of both variable and constant velocity models is  good: the relatively low $R^2$ is a result of the scatter of the many data points for larger times. In this case both models show similar results and a much smaller difference in predicted $k_a$ values than with the data of \cite{Delgado2006}.

\begin{table}[H]
    \centering
    \caption{Results obtained from  fitting to the experimental data of Monazam \emph{et al.} \cite{Monazam2013} at 40$^\circ$C and 70$^\circ$C. The only fitting parameter is $k_a$, the rest are obtained using the values in Table~\ref{tab:Monproperties}.}
    \label{tab:Monresults}
    \begin{tabular}{|c|c|c|c|}
    \hline
      \cellcolor[rgb]{0.7922,0.8471, 0.8706}\textbf{Parameter} & \multicolumn{3}{c|}{\cellcolor[rgb]{0.7922,0.8471, 0.8706}\textbf{Value}}  \\ \hline \hline
      $\phi_1$ (-) & \multicolumn{2}{c|}{$0.333$} & $0.166$ \\ \hline
      $T$ ($^\circ$C) & $40$ & \multicolumn{2}{c|}{$70$} \\ \hline
      Partial orders & (1,2) & \multicolumn{2}{c|}{(1,1)} \\ \hline \hline
      $k_a$ ($\times 10^{-4}$) (m$^3$kg$^{n-1}$mol$^{-n}$s$^{-1}$) & 5.388 & 6.230 & 7.088 \\ \hline
      $k_d$ ($\times 10^{-4}$) (kg$^{n-1}$mol$^{1-n}$s$^{-1}$) & 2.532 & 3.930 & 4.471 \\ \hline
      $\mathcal{T}$ (s) & 60.43 & 135.71 & 239.29 \\ \hline
      $\mathcal{L}$ ($\times 10^{-2}$) (m) & 1.486 & 2.265 & 1.907 \\ \hline
      $\mathcal{P}$ (Pa) & 896.47 & 1246.81 & 1049.69 \\ \hline\hline
      $\delta_1$ ($\times 10^{-3}$) & 2.60 & 1.94 & 0.92 \\ \hline
      $\delta_2$ ($\times 10^{-2}$) & 3.420 & 2.459 & 2.921 \\ \hline
      $\delta_3$ ($\times 10^{-2}$) & 0.885 & 1.231 & 1.036 \\ \hline
      $\delta_4$ ($\times 10^{-3}$) & 2.925 & 2.436 & 3.046 \\ \hline
      $\delta_5$ & 0.784 & 0.784 & 0.313 \\ \hline
      $\delta_6$ ($\times 10^{-6}$) & 5.673 & 3.100 & 1.836 \\ \hline
      $\delta_7$ & 0.036 & 0.053 & 0.107 \\ 
      \hline
    \end{tabular}
\end{table}

The results of the fitting and calculated parameters and scales  are shown in Table~\ref{tab:Monresults}. The value of $k_d$ has been calculated as $k_d=k_a/K$. Parameters $\delta_1$ and $\delta_4$ are both small, with an order of magnitude of 10$^{-3}$. In ths example adsorption strongly dominates and $\delta_7$ is small. Note that $\delta_2$, $\delta_3$ and $\delta_6$ are all less than 0.1, which validates the assumptions made when deriving the analytical solutions from section \ref{sec:AnalyticalSolns}. 
%expressions \eqref{BC11} and \eqref{BC12}.

%At high temperature (70$^\circ$C), Monazam \emph{et al.} \cite{Monazam2013} report breakthrough curves with two different volumetric percentages of CO$_2$: 16.6\% and 33.3\% (5.90 and 11.83 mol/m$^3$). The fitting of the different models compared in Table~\ref{tab:Monresultsgof} at this temperature is shown in \ref{fig:Monc0}. Since the ratio $m/n=1$ is the one that shows a better fit in the isotherm study at high temperatures (see Supplementary Material), only the variable and constant velocity models with $(m,n)=(1,1)$ are compared. 

%As suggested by the SSE and R-squared values in Table~\ref{tab:Monresultsgof}, the fitting with the $(1,1)$ model \eqref{BC11} is excellent. Note that the differences in the fitting between the variable and the constant velocity models (solid and dashed lines, respectively) is scarce with 16.6\% of CO$_2$. This difference is more evident though when the percentage of CO$_2$ increases to 33.3\%. With this CO$_2$ content, the solid line is capturing better the first slope of the breakthrough curve when $c/c_{10}\rightarrow 0$. This can also be seen in Table~\ref{tab:Monresultsgof}, since the SSE is almost the same for 16.6\% of CO$_2$ (27.412 and 31.126), while for 33.3\% of CO$_2$ the variable velocity SSE is 1.297 and the constant velocity is 2.882. 

%\textcolor{blue}{The value of $k_a$ decreases by a factor of 0.88 when $1-\phi_1$ decreases by a factor 0.8. This trend is consistent with the results obtained from Delgado \emph{et al.} \cite{Delgado2006}, which would suggest a clear correlation between $k_a$ and $u$ (and hence with $1-\phi_1$). also for the $(1,1)$ model \eqref{BC11}}. 

We observe that, for the data at 70$^\circ$C in Table~\ref{tab:Monresults} where we have two sets of results,  the value of $k_a$ decreases by a factor of 0.88 when $1-\phi_1$ decreases by a factor 0.8, suggesting an almost linear correlation between $k_a$ and $1-\phi_1$ and hence with $u$. This trend is consistent with our previous analysis of Delgado's data. 

 %%%%%%%%%%%%%%%%%%%%%%%%%%%%%%%%%%%%%%%%%%%%%%%%%%%%%%%%%%%%%%%%%%%%%%%%%%%%%%%%%%%%%%%%

\section{Conclusion}\label{sec:conclusion}

In this paper we have developed a mathematical model that describes the removal of large quantities of contaminant from a fluid mixture via adsorption. Comparison of numerical and approximate analytical solutions demonstrated the accuracy of the approximate solutions. Comparison of the approximate solutions with experimental data demonstrated the accuracy of the method. Specifically we examined the capture of CO$_2$ with volume fractions ranging from 14 to 69\%, finding excellent agreement in all cases.

The key result of the paper is the set of analytical solutions describing the concentration, pressure, velocity and amount adsorbed throughout the column. All of these solutions are novel and represent a
breakthrough in the understanding of the adsorption of large quantities of a fluid. The non-dimensional solution indicated that 
the most important parameters affecting  the breakthrough form are the ratio of adsorption to desorption, the feed contaminant concentration and volume fraction. The dimensional solution shows how this shape is stretched in the time domain, which then adds the adsorption rate, maximum possible adsorption amount and form of chemical reaction to the list of controlling parameters.

Expressions for the breakthrough curve were provided in an implicit form
\bea
t = t_{1/2} + m Y(c) \, ,
\eea
where the form of $Y$ depends on the chemical reaction and volume fraction. Here we derived solutions for the case of physisorption or a 1, 1 chemical reaction (1 contaminant molecule reacts with 1 adsorbent) and also a 1, 2 reaction. More complicated reactions could be described following the same solution method. The  standard route is to describe breakthrough concentration as a function of time, via the explicit form $c = f(t)$. Plotting the former, as a straight line graph, $t$ versus $Y(c)$, against experimental data provides a  clear picture of which model (i.e. which chemical reaction, large mass  or trace removal) is applicable. Plotting results in the explicit form, the differences between model and data are not always clear, making it easier to accept an incorrect model. This assertion was confirmed through two groups of experimental data, in both cases good agreement could be obtained between trace amount of contaminant result and the large mass one when plotting $c$ as a function of time. However, when using the linear form it was clear that the trace amount model did not correctly predict the behaviour and so could not be trusted.

Existing models of trace removal rely on 
the adsorption coefficient, $k_a$,  remaining constant with respect to concentration. For example, the Langmuir mass loss term has $\partial q/\partial t \propto k_a c (q_{max} - q)$. To obtain an analytical solution integration is carried out assuming $c$, $q$ are variable and $k_a$ is constant. If solutions show that $k_a=k_a(c)$ then the integrations are invalid and the solution no longer holds. Instead the kinetic equation should be rewritten, for example with $k_a c^m$, such that $k_a$ is independent of $c$, and a new integration carried out. However, it is well-known that the adsorption rate may vary with pressure and flow rate. Removing trace amounts these do not vary significantly, removing large amounts both may vary. Consequently we observed a decrease in $k_a$ with increasing volume fraction of contaminant (corresponding to a reduction in velocity and  pressure with respect to the trace amount model).

Future work will concentrate on expanding the solutions to different chemical reactions, whenever appropriate experimental data is available to verify the results as well as investigating the relation between velocity and adsorption coefficient.  While we have focussed on parameter regimes relevant to adsorption columns, so motivating model reductions, the  models may apply to a variety of sorption phenomena with different parameters, different reductions and different behaviour. This could be a rich vein for subsequent investigations.

%%%%%%%%%%%%%%%%%%%%%%%%%%%%%%%%%%%%%%%%

\section*{Declaration of Competing Interest}
The authors declare that they have no known competing financial interests or personal relationships that could have appeared to
influence the work reported in this paper.

%%%%%%%%%%%%%%%%%%%%%%%%%%%%%%%%%%%%%%%%%%%%%%%%%%%%%%%%%%%%%%%%%%%%%%%%%%%%%%%%%%%%%%%%

\section*{Acknowledgements}
This publication is part of the research projects PID2020-115023RB-I00 and TED2021-131455A-I00 (funding T. Myers and F. Font) financed by
MCIN/AEI/ 10.13039/501100011033/, by “ERDF A way of making Europe” and by “European Union NextGenerationEU/PRTR”. 
A. Valverde acknowledges support from the Margarita Salas UPC postdoctoral grants funded by the Spanish Ministry of Universities with European Union funds - NextGenerationEU. 
T. Myers and M. Calvo-Schwarzwalder thank CERCA Programme/Generalitat de Catalunya for institutional support. 
This work is supported by the Spanish State Research Agency, through the Severo Ochoa and Maria de Maeztu Program for Centres and Units of Excellence in R\&D (CEX2020-001084-M). 
F. Font is a Serra-Hunter fellow from the Serra-Hunter Programme of the Generalitat de Catalunya. F. Font gratefully acknowledges the SRG programme (2021-SGR-01045) of the Generalitat de Catalunya (Spain).

%%%%%%%%%%%%%%%%%%%%%%%%%%%%%%%%%%%%%%%%%%%%%%%%%%%%%%%%%%%%%%%%%%%%%%%%%%%%%%%%%%%%%%%%

%\bibliographystyle{elsarticle-num}
\bibliography{bibcarbon}

\end{document}

\maketitle

\section{Analytical approximation}\label{SM:AnalyApprox}

\subsection{Reduced model}\label{SM:model}
Neglecting all terms deemed small through the non-dimensionalisation, the governing equations are
\begin{subequations}\label{SM:governingeqs}
\begin{align}
    \pad{}{\hat x}\left(\hat u\hat c_1\right)&=-\pad{\hat q}{\hat t}\, ,\label{SM:ndc1eq}\\
    \pad{}{\hat x}\left(\hat u\hat c_2\right)&=0\, ,\label{SM:ndc2eq}\\
    1& =    \phi_1\hat c_1 +(1-\phi_1)\hat c_2\, ,\label{SM:ndpeq}\\
    -\pad {\hat p}{ \hat x} &= \hat u\, ,\label{SM:ndpxeq}\\
    \pad{\hat q}{\hat t}&=\hat c_1^m(1-\hat q)^n-\delta_7\hat q^n\, .\label{SM:ndqeq}
\end{align}
\end{subequations}
where $\phi_1$ is the contaminant volume fraction and $\delta_7$ is the adsorption-to-desorption ratio. The boundary and initial conditions are 
\begin{subequations}\label{SM:nondimBCs}
\begin{align}
    &1=   \hat u(0,\hat t)\hat c_1(0,\hat t) \, ,\qquad \left.\pad{\hat c_1}{\hat x}\right|_{\hat x=\hat L}=0\, ,\qquad \hat c_1(\hat x,0)=0\, ,\label{SM:nondimBCs1}\\
    &1 =  \hat u(0,\hat t)\hat c_2(0,\hat t) \, ,\qquad \left.\pad{\hat c_2}{\hat x}\right|_{\hat x=\hat L}=0\, ,\qquad \hat c_2(\hat x,0)=1/(1-\phi_1)\, ,\label{SM:nondimBCs2}\\
    &
    \hat p(0,\hat t) = \Delta\hat p\, ,\quad\quad\quad\quad 
    \hat p(\hat L,\hat t)=0\, ,\quad\quad \hat p(\hat x,0)=\hat p_{in}(\hat x)\, ,\label{SM:nondimBCs3}\\  
    &\hat q(\hat x,0)=0\, ,\label{SM:nondimBCs4}
\end{align}
\end{subequations}
with $\hat p_{in}(\hat x)=\Delta\hat p(0)\left(1-{\hat x}/{\hat L}\right)$. From this reduced formulation we can immediately determine $\hat c_2$, $\hat u$, and $\hat p$ in terms of $\hat c_1$,
\begin{equation}\label{SM:reduscedsolns}
    \hat c_2 = \frac{1-\phi_1\hat c_1}{1-\phi_1}\, ,\qquad \hat u = \frac{1}{\hat c_2} = \frac{1-\phi_1}{1-\phi_1\hat c_1}\, ,\qquad \hat p=\int_{\hat x} ^{\hat L} \hat u(\hat\xi,\hat t)d\hat\xi\, ,
\end{equation}
which reduces the problem to determining $\hat c_1$ and $\hat q$. Upon substituting $\hat u$ in Eq. \eqref{SM:ndc1eq} the equation for $\hat c_1$ becomes
\begin{align}\label{SM:ndc1eqred}
    \pad{}{\hat x}\left(\frac{(1-\phi_1)\hat c_1}{1-\phi_1\hat c_1}\right)&=-\pad{\hat q}{\hat t}\, ,
\end{align}
which is coupled to the kinetic equation \eqref{SM:ndqeq}.

\subsection{Calculation of travelling wave solutions}\label{app:TWsolns}
The reduced system formed by Eqs. \eqref{SM:ndqeq} and \eqref{SM:ndc1eqred} can be simplified using a common approach based on introducing a travelling wave coordinate. For this we define $\hat \eta=\hat x-\hat s(\hat t)$, where $\ud\hat s/\ud\hat t=\hat v>0$, and define 
\begin{equation}
    \frac{(1-\phi_1)\hat c_1}{1-\phi_1\hat c_1}=:\hat F(\hat\eta)\, ,\qquad \hat q(\hat x,\hat t)=: \hat G(\hat \eta)\, .
\end{equation}
Equation \eqref{SM:ndc1eqred} becomes
\begin{equation}
    \nd{\hat F}{\hat\eta}=\hat v\nd{\hat G}{\hat\eta}\, ,
\end{equation}
which can be directly integrated to 
\begin{equation}\label{TW_F}
    \hat F=\hat v\hat G\, ,
\end{equation}
where the constant of integration has been chosen to be zero to satisfy $\hat F,\hat G\to0$ as $\hat\eta\to\infty$, i.e., that no contaminant is present far ahead of the moving front. The kinetic equation \eqref{SM:ndqeq} becomes
\begin{equation}\label{Anal_G}
    -\hat v\nd{\hat G}{\hat\eta}=\left(\frac{\hat F}{1-\phi_1+\phi_1\hat F}\right)^m(1-\hat G)^n-\delta_7\hat G^n\, .
\end{equation}

Far behind the moving front, the process is expected to have reached an equilibrium state and therefore $\hat F\to\hat F_e$ and $\hat G\to\hat G_e$ as $\hat\eta\to-\infty$. Moreover, we expect $\hat c_1\to1$ far behind the moving front where the system is in equilibrium. Using this in the definition of $\hat F$, we find 
\begin{equation}\label{def:Fe}
    \hat F_e=1\, ,
\end{equation}
whereas  taking the limit as $\hat\eta\to-\infty$ in Eq. \eqref{Anal_G} with $\ud\hat G/\ud\hat\eta\to0$ yields 
\begin{equation}\label{def:Ge}
    \hat G_e=\frac{1}{1+\delta_7^{1/n}}\, ,
\end{equation}
which is  the non-dimensional equilibrium value of $\hat q$ (as specified in the main text). Taking the limit as $\hat\eta\to-\infty$ in Eq. \eqref{TW_F} determines the wave velocity
\begin{equation}\label{TW:def:v} 
    \hat v=\frac{\hat F_e}{\hat G_e}=1+\delta_7^{1/n}\, .
\end{equation}

Finally, it is left to solve Eq. \eqref{Anal_G}, which can be combined with Eq. \eqref{TW_F} to give a single ordinary differential equation for $\hat F$,
\begin{equation}\label{TW_final}
    -\nd{\hat F}{\hat \eta}=\left(\frac{\hat F}{1-\phi_1+\phi_1\hat F}\right)^m\left(1-\frac{\hat F}{\hat v}\right)^n-\delta_7\left(\frac{\hat F}{\hat v}\right)^n\, ,
\end{equation}
This equation must be solved numerically in general, but we can find  analytical solutions for certain combinations of $n$ and $m$ which are of interest.
As for where the front is defined, we define $\hat\eta=0$ as the point where $\hat c_1=1/2$, hence  
\begin{equation}\label{def:F12}
    \hat F(0)=\hat F_{1/2}:=\frac{1-\phi_1}{2-\phi_1}\, .
\end{equation}
Upon defining the origin of the wave variable $\hat\eta$ in this way, we can write it explicitly as $\hat\eta=\hat x-\hat L-\hat v\left(\hat t-\hat t_{1/2}\right)$, where $t_{1/2}$ is the (experimentally obtained) time when the outlet concentration is half the inlet value and $\hat t_{1/2}=t_{1/2}/\mathcal{T}$.

%%%%%%%%%%%%%%%%%%%%%%%%%%%%%%%%%%%%%%%%%%%%%%%%%%%%%%%%%%%%%%%%%%%%%%%%%%%%%%%%%%%%%%%%

\subsubsection{Case $m=1,n=1$}\label{ssec:case11}

In this case, Eq. \eqref{TW_final} may be written
\begin{align}\label{TW_11}
-   \nd{\hat{F}}{{\hat\eta}} & = \frac{\hat F}{1-\phi_1 + \phi_{1}\hat F} \left(1 - \frac{\hat F}{\hat v}\right) -\delta_7 \frac{\hat F}{\hat v} = \frac{a\hat F (1-\hat F)}{1-\phi_1 + \phi_{1}\hat F}\, ,
\end{align}
where $a=(1+\phi_1\delta_7)/\hat v > 0$. Rearranging and splitting the fraction yields
\begin{align}
    -1=\left[\frac{1-\phi_1 + \phi_{1}\hat F}{a\hat F (1-\hat F)}\right]\pad{\hat F}{\hat\eta}=\left[\frac{1-\phi_1}{a\hat F}+\frac{1}{a(1-\hat F)}\right]\nd{\hat F}{\hat\eta}\, .
\end{align}
This equation can be directly integrated to give
\begin{equation}
    \mathcal{C}-\hat\eta=\frac{1-\phi_1}{a}\ln\left({\hat F}\right)
    -\frac{1}{a}\ln\left({1-\hat F}\right)\, .
\end{equation}
applying $\hat F(0)=\hat F_{1/2}$ determines the constant of integration to give the solution
\begin{equation}\label{SM:soln:n1}
    -\hat\eta =\frac{1-\phi_1}{a}\ln\left(\frac{\hat F}{\hat F_{1/2}}\right)-\frac{1}{a}\ln\left(\frac{1-\hat F}{1-\hat F_{1/2}}\right)\, .
\end{equation}

%%%%%%%%%%%%%%%%%%%%%%%%%%%%%%%%%%%%%%%%%%%%%%%%%%%%%%%%%%%%%%%%%%%%%%%%%%%%%%%%%%%%%%%%

\subsubsection{Case $m=1,n=2$}\label{ssec:case12}

Equation \eqref{TW_final} now can be rearranged to give
\begin{equation}\label{TW_12}
    -\nd{\hat F}{\hat \eta}=\frac{\hat F}{1-\phi_1+\phi_{1}\hat F}\left(1-\frac{\hat F}{\hat v}\right)^2-\delta_7\left(\frac{\hat F}{\hat v}\right)^2=\frac{\hat F\left(1-\hat F\right)\left(a-\hat F\right)}{a(1-\phi_1+\phi_{1}\hat F)}\, ,
\end{equation}
with $a=\hat v^2/(1-\phi_1\delta_7) =(1+\sqrt{\delta_7})^2/(1-\phi_1\delta_7)$. 
We may rewrite this as 
\begin{align}\begin{split}
    -\frac{d\hat\eta}{a} 
 = \left[\frac{1-\phi_1+\phi_{1}\hat F}{\hat F\left(1-\hat F\right)\left(a-\hat F\right)}\right] d\hat F\, .
\end{split}\end{align}
The value $a \ge 1$ for $\delta_7 \in [0,1/\phi_1)$ and $a \le -1/\phi_1$ for $\delta_7 > 1/\phi_1$.

\paragraph{Case i) $a= 1+\omega^2 \ge 1$.}
 Writing $a$ in terms of $\omega^2 >0$, the integration is simple. After replacing $\omega^2$ and rearranging we obtain
\begin{equation}\label{SM:soln:n2}
    -\hat\eta=(1-\phi_1)\ln\left(\frac{\hat F}{\hat F_{1/2}}\right) - \frac{a}{a-1}\ln\left(\frac{1-\hat F}{1-\hat F_{1/2}}\right) + \left(\phi_1+\frac{1}{a-1}\right)\ln\left(\frac{a-\hat F}{a-\hat F_{1/2}}\right)\, ,
\end{equation}
where we have used $\hat F(0)=\hat F_{1/2}$.

\paragraph{Case ii) $a= -1/\phi_1 -\omega^2 = -\nu^2 < 0$.}

Writing $a$ in terms of $-\nu^2$, we may follow the same route as for the previous calculation and obtain the same result, but note the final log term is now valid since both denominator and numerator are negative (for the case $a \ge 1$ they are both positive).

%%%%%%%%%%%%%%%%%%%%%%%%%%%%%%%%%%%%%%%%%%%%%%%%%%%%%%%%%%%%%%%%%%%%%%%%%%%%%%%%%%%%%%%%

\subsection{Final solutions}
Equation \eqref{TW_F} provides a simple relation between  $\hat F,\hat G$. In terms of the original non-dimensional variables, $\hat c_1,\hat q$, this leads to
\begin{equation}
    \hat q = \frac{(1-\phi_1)\hat c_1}{\left(1+\delta_7^{1/n}\right)\left(1-\phi_1\hat c_1\right)}\, .
\end{equation}
Writing $\eta$ in terms of the original variables
 $\hat c_1$ is defined implicitly by 
\begin{equation}
    \left(1+\delta_7^{1/n}\right)\left(\hat t-\hat t_{1/2}\right)-\left(\hat x-\hat L\right)=\hat Y_{mn}\left(\hat c_1\right)\, ,
\end{equation}
where the expression of $\hat Y_{mn}$ come from the right hand sides of either Eqs. (\ref{SM:soln:n1}, \ref{SM:soln:n2}). Rearranging and replacing $F$ the final expressions are 
\begin{subequations}\label{SM:TWsolution:Yc}
\begin{align}
    \hat Y_{11}(\hat c_1) &= \frac{1+\delta_7}{1+\phi_1\delta_7}\left[\ln\left|\frac{\hat c_1}{1-\hat c_1}\right|-\phi_1\ln\left|\frac{(2-\phi)\hat c_1}{1-\phi_1\hat c_1}\right|\right]\, ,\\
    \hat Y_{12}(\hat c_1) &= \frac{1+\sqrt{\delta_7}}{\sqrt{\delta_7}\left(1+\sqrt{\gamma}\right)}\left[\ln\left|\frac{\hat c_1}{1-\hat c_1}\right|-\gamma\ln\left|\frac{(2-\gamma)\hat c_1}{1-\gamma\hat c_1}\right|\right]\, ,
\end{align}
and
\begin{equation}
\gamma=\left(\frac{1+\phi_{1}\sqrt{\delta_7}}{1+\sqrt{\delta_7}}\right)^2 \, . \label{SM:eq:gamma}
\end{equation}\end{subequations}
The remaining quantities are determined by Eq. \eqref{SM:reduscedsolns}.

\section{Isotherm study of CO$_2$ capture}\label{SM:IsoAppCO2}

\subsection{CO$_2$ adsorption onto silicate pellets from helium stream}\label{sec:DelIso}

\begin{figure}[H]
  \centering
  \includegraphics[width=0.48\textwidth]{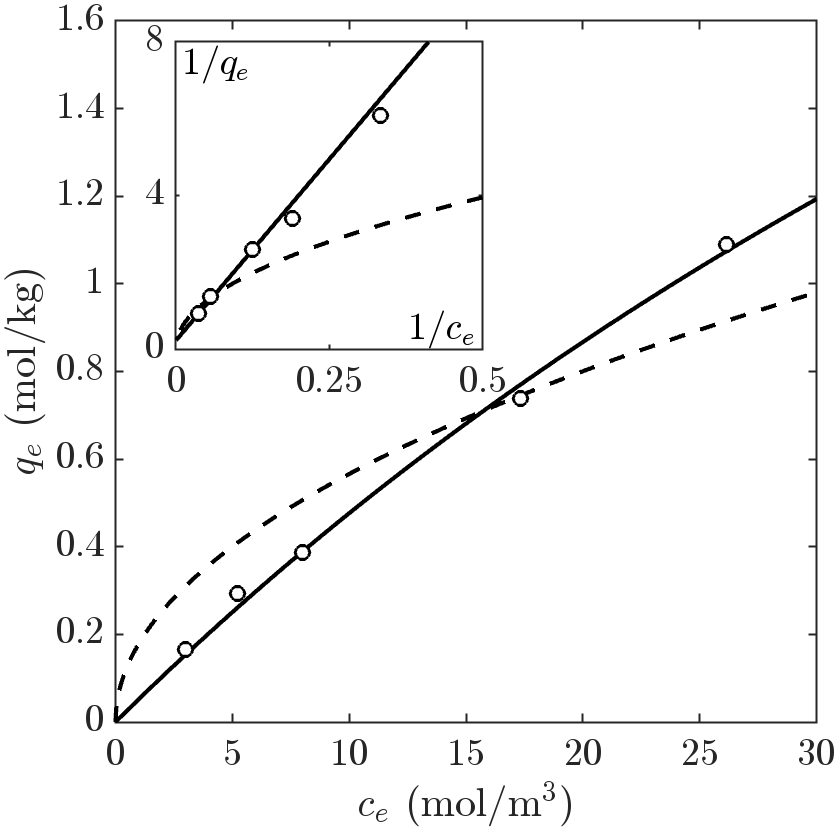}
  \caption{Fitting of the experimental isotherm data reported by Delgado \emph{et al.} \cite{Delgado2006}. The solid line refers to $m/n=1$ and the dashed line to $m/n=0.5$. $c_{e}$ refers to the inlet concentration at which equilibrium is reached in the column. Insets show $1/c_e$ vs $1/q_e$. As these isotherms are obtained via column adsorption experiments, we have $c_e=c_{10}$.}
  \label{fig:Delgadoisotherm}
\end{figure}

Delgado \emph{et al.} \cite{Delgado2006}  provide the isotherm obtained during their column experiments. The data points and fits using the Sips isotherm with ratios $m/n=1$ and $m/n=0.5$ are shown in Fig.~\ref{fig:Delgadoisotherm}. The parameters $q_{max}, K$ obtained from the fitting of the isotherm are provided in Table~\ref{tab:Tableiso}. These values are  used in the main document so that only one unknown remains in the fitting of the breakthrough data.

\begin{table}[H]
\centering
\caption{Values of the isotherm parameters obtained from fitting the Sips isotherm to the experimental data reported by Delgado \textit{et al.} \cite{Delgado2006}. }
\label{tab:Tableiso}
\begin{tabular}{|c|c|c|c|c|}
\hline
Orders &  $q_{max}$ & $K$ & SSE & \multirow{2}{*}{R-squared}\\
ratio &  (mol/kg) & (m$^{3m}$kg$^{n-1}$mol$^{1-m-n}$) & (mol$^2$/kg$^2$) & \\ 
\hline \hline
$m/n=0.5$ & 7.0$\times$10$^{11}$ & 2.6$\times$10$^{-13}$ & 0.0793 & 0.8598 \\ \hline
$m/n=1$  & 4.8224 & 0.01097 & 0.0024 & 0.9957 \\ \hline
\end{tabular}
\end{table}

The ratio $m/n=1$ shows  excellent agreement with the isotherm data, whereas the ratio $m/n=0.5$ clearly shows a poor fit. This is consistent with the adsorption mechanism of CO$_2$ on silicalite pellets. Near ambient temperature CO$_2$ molecules are adsorbed by Van der Waals forces inside the micropores of the silicalite structure, mainly by dispersion forces \cite{Grajciar2012,Jiang2020,Santoro2011}. The Langmuir isotherm ($m/n=1$ ratio) accounts for  homogeneous adsorption of isolated molecules in the most favoured sites of the silicalite structure \cite{Lill2022}. Consequently, in the main document we only employ  $(m,n)=(1,1)$ using the associated $q_{max}, K$ values.

\subsection{CO$_2$ capture onto PEI modified silica from N$_2$ stream}\label{sec:MonIso}

\begin{figure}[H]
  \centering
  \begin{subfigure}[b]{0.48\textwidth}
         \centering
         \includegraphics[width=\textwidth]{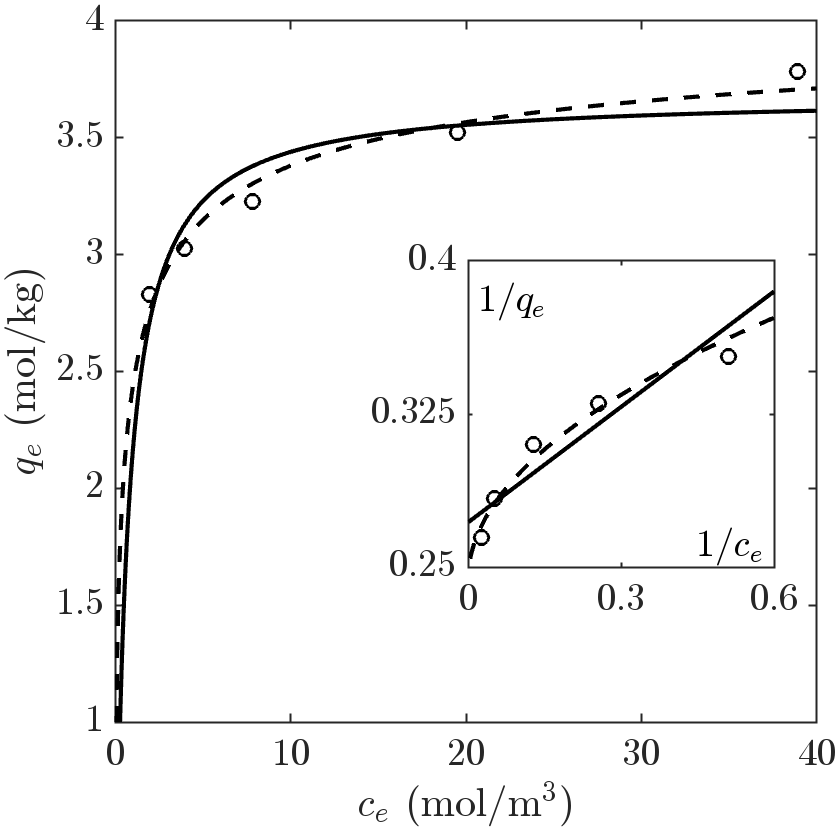}
         \caption{$T=40$ $^\circ$C.}
         \label{fig:Monisotherms40}
     \end{subfigure}
     \hfill
     \begin{subfigure}[b]{0.48\textwidth}
         \centering
         \includegraphics[width=\textwidth]{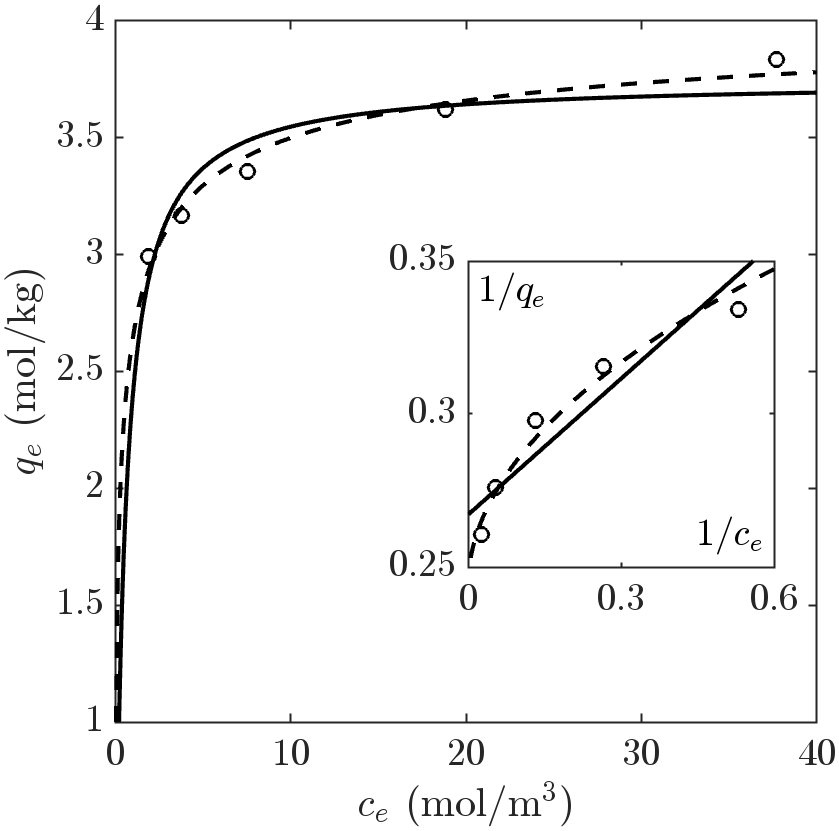}
         \caption{$T=50$ $^\circ$C.}
         \label{fig:Monisotherms50}
     \end{subfigure}
     \\
     \begin{subfigure}[b]{0.48\textwidth}
         \centering
         \includegraphics[width=\textwidth]{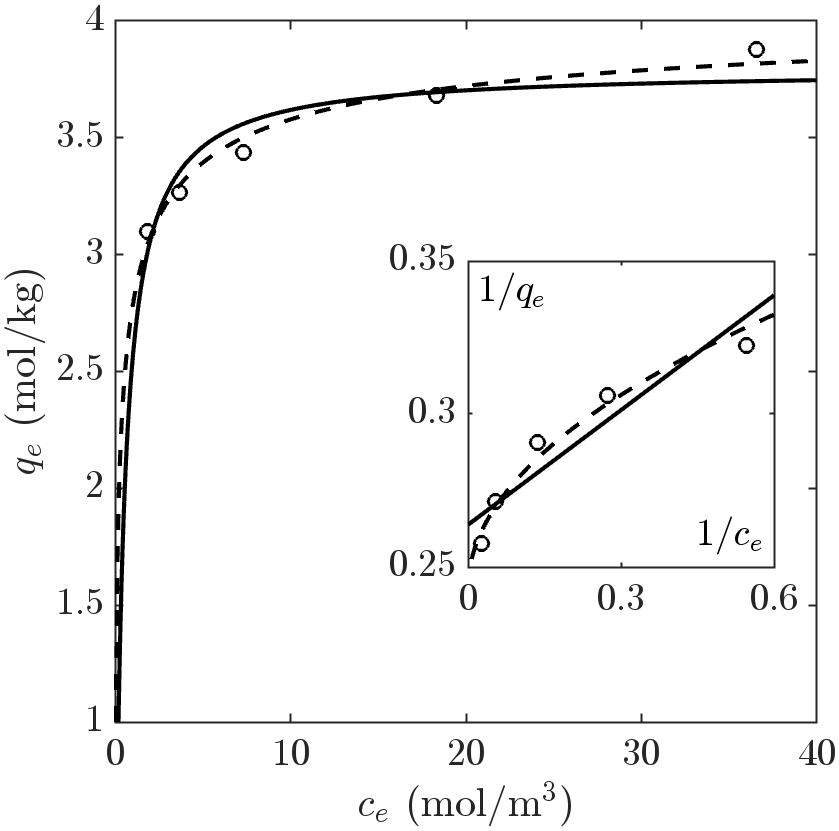}
         \caption{$T=60$ $^\circ$C.}
         \label{fig:Monisotherms60}
     \end{subfigure}
     \hfill
     \begin{subfigure}[b]{0.48\textwidth}
         \centering
         \includegraphics[width=\textwidth]{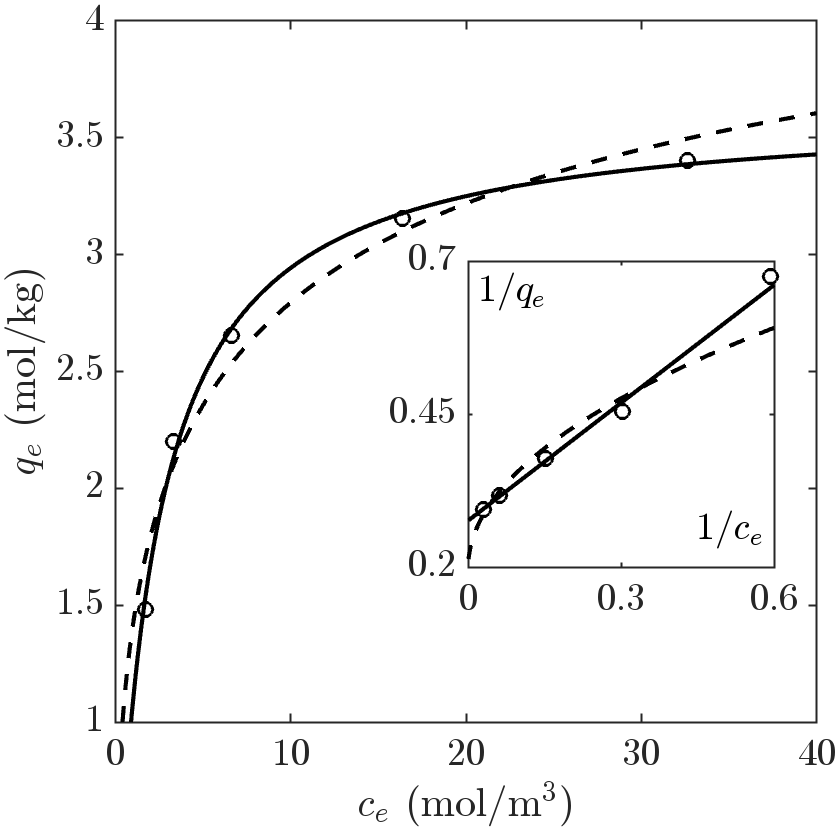}
         \caption{$T=100$ $^\circ$C.}
         \label{fig:Monisotherms100}
     \end{subfigure}
  \caption{Fitting of the experimental isotherms reported by Monazam \textit{et al.} \cite{Monazam2013iso} using the Sips isotherm. The solid line refers to $m/n=1$ (Langmuir isotherm) and the dashed line to $m/n=0.5$. The symbol $c_e$ refers to the value of the concentration when equilibrium is reached inside the batch reactor. Insets show $1/c_e$ vs $1/q_e$.}
  \label{fig:Monisotherms}
\end{figure}

In Monazam \emph{et al.} \cite{Monazam2013iso} equilibrium (batch) data is provided for CO$_2$ capture on PEI modified silica from N$_2$. This is the same adsorbate-adsorbent system as studied in the column adsorption experiments of Monazam \emph{et al.} \cite{Monazam2013}. In \cite{Monazam2013iso}, only the isotherms at 40$^\circ$C and 100$^\circ$C are provided. However, they also provide the coefficients obtained from fitting the isotherms at 50$^\circ$C, 60$^\circ$C and 80$^\circ$C with Freundlich and Langmuir isotherms. Since the fits with the Freundlich isotherm at 50$^\circ$C and 60$^\circ$C show good agreement (R-squared 0.992 and 0.989, respectively), we created data points for these temperatures using Freundlich's coefficients in order to carry out a new fitting with the Sips isotherm accounting for $m/n=1$ and $m/n=0.5$. The results are shown in Figure \ref{fig:Monisotherms}.

The parameters obtained from the fitting of each isotherm  are reported in Table~\ref{tab:TableisoMon}.

\begin{table}[H]
\centering
\caption{Parameters values obtained from fitting the Sips isotherm to the experimental batch data of Monazam \textit{et al.} \cite{Monazam2013iso}. The subscript $b$ in $q_{max,b}$ refers batch.}
\label{tab:TableisoMon}
\begin{tabular}{|c|c|c|c|c|}
\hline
$T$ ($^\circ$C)&  40 & 50 & 60 & 100\\
\hline \hline
\multicolumn{5}{|c|}{\textbf{$m=1$ $n=2$}} \\
\hline
$q_{max,b}$ (mol/kg)  & 4.114 & 4.113 & 4.119 & 5.077 \\ \hline
$K$ (m$^3$kg/mol$^2$)  & 2.128 & 3.275 & 4.400 & 0.150 \\ \hline
SSE (mol$^2$/kg$^2$)  & 0.0189 & 0.0138 & 0.0122 & 0.0846 \\ \hline
R-squared  & 0.9675 & 0.9701 & 0.9689 & 0.9642 \\
\hline\hline
\multicolumn{5}{|c|}{\textbf{$m=1$ $n=1$}} \\
\hline
$q_{max,b}$ (mol/kg) & 3.679 & 3.745 & 3.791 & 3.629 \\ \hline
$K$ (m$^3$/mol) & 1.440 & 1.801 & 2.111 & 0.429 \\ \hline
SSE (mol$^2$/kg$^2$)  & 0.0752 & 0.0576 & 0.0496 & 0.0085 \\ \hline
R-squared  & 0.8708 & 0.8751 & 0.8731 & 0.9964 \\
\hline
\end{tabular}
\end{table}

Both Fig.~\ref{fig:Monisotherms} and the results in Table~\ref{tab:TableisoMon} show that the Sips isotherm with $m/n=0.5$ fits the experimental data up to 60$^\circ$C, whereas for 100$^\circ$C the better fit occurs  with $m/n=1$. 
The equilibrium constant $K$ increases from 40$^\circ$C to 60$^\circ$C, while $q_{max,b}$ remains approximately constant. The increase of $K$  results in a steeper isotherm, which means that for a fixed $c_e$ the equilibrium value $q_e$ will also increase with temperature.  However, there is a change in behaviour observable in the  100$^\circ$C result, which shows a far lower $K$value (and consequently lower $q_e$ values). In physisorption processes $K$ is expected to decrease with temperature, since the intermolecular forces responsible for retaining the adsorbate molecules,  weaken with increasing temperature. However, chemisorption processes may show an opposite trend when the enthalpy of reaction is positive.

In the main text we analyse breakthrough data at 70$^\circ$C. 
The value of the equilibrium constant used in this case was calculated by extrapolation of the results at 80$^\circ$C reported by \cite{Monazam2013iso}, and the values obtained at 100$^\circ$C from Fig.~\ref{fig:Monisotherms} using the Van't Hoff equation \cite{Auton2024,Atkins2006}. 

The adsorbed fractions may differ between batch and column experiments, however, we expect the equilibrium constant $K$ to be relatively constant between the two. Consequently we calculate $q_{max}$ for column experiments using the Sips isotherm and substituting for $K$ from the batch results and $q_e$
from the column experiment.

\bibliographystyle{unsrt}
\bibliography{bibcarbon}